\documentclass[11pt, oneside]{article}   	
\usepackage{geometry} 
\usepackage{color}               		
\usepackage{graphicx}				
\usepackage{amssymb}
\usepackage{setspace}

\usepackage{appendix}   

\usepackage{authblk}

\usepackage[semicolon]{natbib}
\usepackage{verbatim}
\usepackage{soul}
\usepackage{graphicx}
\usepackage{subcaption}
\usepackage{hyperref}

\usepackage{tikz}	
\usetikzlibrary{shapes, arrows}
\tikzstyle{compartment} = [circle, minimum width=2cm, text centered, draw=black, fill=none]
\tikzstyle{arrow} = [thick,->,>=stealth]

\usepackage{multirow}

\newcommand{\SARS}{SARS-CoV-2}

\newcommand{\nInfect}{10} 

\newcommand{\bs}{\textbf{S}}
\newcommand{\be}{\textbf{E}}
\newcommand{\ba}{\textbf{A}}
\newcommand{\bia}{\textbf{I1}}
\newcommand{\bib}{\textbf{I2}}
\newcommand{\bi}{\textbf{I}}
\newcommand{\br}{\textbf{R}}

\newcommand{\numDays}{90}	
\newcommand{\numInitialCases}{10}	
\newcommand{\numReps}{10}	
\newcommand{\numParCombs}{26,244}

\title{\SARS\ Transmission in University Classes}
\author[1]{William Ruth}
\author[2]{Richard Lockhart}
\affil[1]{Corresponding Author - Department of Statistics and Actuarial Science \\ Simon Fraser University \\ Burnaby, BC  Canada \\ wruth@sfu.ca}
\affil[2]{Department of Statistics and Actuarial Science \\ Simon Fraser University \\ Burnaby, BC  Canada}
\date{}

\begin{document}
\maketitle

\doublespacing

\begin{abstract}
	We investigate transmission dynamics for \SARS\ on a real network of classes at Simon Fraser University, a medium-sized school in Western Canada. Outbreaks are simulated over the course of one semester across numerous parameter settings for a realistic compartment model, including asymptomatic and presymptomatic transmission. We investigate the control strategy of moving large classes online while small classes are allowed to meet in person. Regression trees are used to model the effect of disease parameters on simulation outputs; specifically, the total number of infections and the peak number of simultaneous cases. We find that an aggressive class size thresolding strategy is required to mitigate the risk of a large outbreak, and that transmission by symptomatic individuals is a key driver of outbreak size.
\end{abstract}

\section{Introduction}
\label{introduction}

\subsection{Past Work}

Statistical and mathematical models are powerful tools for studying the \SARS\ pandemic. Many authors have developed sophisticated models to predict the spread of the disease, which have influenced policy and, ultimately, saved lives \citep{Ves20}. The problem of disease modelling is large and multifaceted. Here, we focus exclusively on transmission within a the context of a university. Specifically, we investigate the effect of moving certain classes online, to see whether limited in-person instruction can be maintained while preventing a major outbreak. Our data come from Simon Fraser University in Burnaby, Canada, but the framework can be applied to other institutions.

Giving a complete overview of the \SARS\ modelling literature here would be impossible. We provide only a brief summary of some closely related work. Models can be broadly classified into two categories: individual level, and differential equation based \citep{Bra19}. Individual level models investigate the effects of individual agents' actions on population level outcomes, whereas differential equation based techniques involve directly modelling population level phenomena. We work entirely within the individual level model framework. See \citet{Est20} for an overview of differential equation models for \SARS\ spread. See \citet{Kis17} for a thorough overview of modelling and analysis of disease spread on networks.

Many individual level models have a compartmental structure, such as SIR (susceptible, infectious, removed) or SEIR (susceptible, exposed, infectious, removed) \citep{Bra08, Dea10}, where individuals are assigned to a category based on their disease status, and the researcher models how individuals move between categories. While much work has been done on modelling community transmission (see, e.g., \citealp{BC21, Cha20, Rad20, Tui20}), some authors have instead directed their efforts toward understanding outbreaks on university campuses \citep{Gre20, Zho21, Kha20, Bor20, Bah20, Fra20, Amb21, Chr20, Wee20}. \citet{Gre20} simulated social dynamics within a university and the corresponding infection rates. They examined the effects of various interventions, including mask wearing, remote instruction, and random testing; with particular attention paid to the test's false-positive rate (i.e.\ specificity). \citet{Zho21} use simulation to investigate the effects of several control strategies with a simplified model of \SARS\ dynamics. \citet{Kha20} developed a detailed framework for simulating infections, which integrates models of various phenomena related to the disease. \citet{Bor20} studied the effects of different strategies for grouping students in dorms and classes. \citet{Bah20} developed a detailed model of how students and faculty interact on a small university campus. In a seminar presentation, \citet{Fra20} discussed both individual-level and differential equation models for disease spread at a large campus; paying particular attention to universal testing schemes and strategies for contact tracing.  \citet{Amb21} developed a framework for assessing risk of infection over the course of a semester based on room crowding and air circulation. \citet{Chr20} give a rapid review of studies modelling COVID transmission in universities.

\citet{Wee20} took a different approach to investigating enrollment at Cornell University. Instead of studying disease transmission directly, they measured numerous graph-theoretic properties of the enrollment network. Their focus was on measuring the connectedness of the network.

\subsection{Our Contribution}
\label{sec:our_cont}

We received data on enrollments at Simon Fraser University (SFU), a medium-sized school located just outside of Vancouver, Canada. Our dataset contains enrollment records from the fall term in 2019 consisting of 110,000-120,000 entries, where each entry corresponds to a specific course taken by a specific student\footnotemark. We also have records of the days on which each class meets, but not at what time. Our dataset does not include any distance learning courses, co-op courses (a.k.a.\ work experience), or courses that do not meet at one of SFU's main campuses. A number of classes in the dataset do not have any meeting days. 

\footnotetext{We also have data for spring 2019 and 2020 terms, but limit our investigation to fall 2019 for the sake of brevity.}

Along with many other universities, SFU adopted a near-total lockdown policy in response to the \SARS\ virus and has only recently returned to in-person instruction. Although this lockdown dramatically reduced the possibility of on-campus transmission, it has also adversely impacted students' learning. There was also an interim period where some classes were held on campus while others remained virtual. The partial return to campus model has the clear advantage of allowing many classes to meet in-person, but also carries an increased risk of infection. Particularly catastrophic would be an outbreak on campus, where a large proportion of the student body becomes infected.

The goal of our study is to investigate potential outbreaks on SFU's main campus when a limited number of smaller classes are allowed to meet in-person. We focus particularly on how properties of these outbreaks vary as the size of in-person classes varies. Although there are countless ways in which students can infect each other on and off campus, we focus on disease transmission through classes. As such, we omit all classes which do not have a scheduled meeting day. Ideally, we would have investigated these courses further. However, for privacy reasons we do not have identifying information for any of these courses, and are thus unable to learn any more about them. Removing these courses will undoubtedly have changed the structure of the enrollment network at SFU, but not in a way that impacts person-to-person contact and thus disease transmission (ostensibly, there is no in-person interaction in a course with no meeting days). We treat labs and tutorials as distinct classes with no inherent connection to the main course with which they are affiliated (other than overlapping enrollment), since each meeting, be it lecture or tutorial, is a separate opportunity for disease spread.

The enrollment network at SFU contains a number of isolated groups of students. That is, groups of students who share classes with each other, but not with anyone outside the group. In graph theory, these groups are called connected components of the network \citep{Cla91}. Since the only avenue of disease transmission that we study is via shared classes, and there are no shared classes between components, we focus on one connected component at a time. It turns out that each term's network is dominated by a single large component, so we keep only this main component and omit all the small ones. Note that the inclusion or exclusion of tutorials has no effect on which students belong to the same connected component (Alice and Bob share a tutorial if and only if they also share the class with which that tutorial is affiliated).

To model disease spread, we simulate transmission over the course of one term. We start by infecting \nInfect\ randomly chosen individuals. We then track how the disease spreads through classes over 90 days (roughly the duration of the pre-exam portion of a term at SFU), with numerous different regimes for the epidemiological properties of the disease. We also consider several schemes for moving large classes online to slow the spread of the infection. Multiple simulations are run under each regime, then various numerical and graphical summaries are reported.

It is important to note that the results of our simulations should not be taken literally. There are many factors that influence how a disease might spread across a university campus, and we can't hope to model all of them. As such, our findings are meant to be interpreted qualitatively; suggesting trends across variables, rather than as a tool to set specific policy strategies. 

Computation is done using the \texttt{R} \citep{R21} and \texttt{Julia} \citep{Bez17} programming languages.

The rest of this paper proceeds as follows. Section \ref{sec:sim} discusses our disease model and computational framework. Section \ref{sec:analysis} describes the analysis we perform, and Section \ref{sec:results} presents the results. Section \ref{sec:discussion} contains interpretation of our results and some discussion of the limitations of our study. Finally, \ref{sec:conclusions} gives some broader implications of our work.

\section{Simulation Study}
\label{sec:sim}

In order to investigate the relationship between network structure and disease transmission, we carry out a simulation study. We use an \textbf{SEAIR} compartment model for the behaviour of \SARS. Respectively, these compartments correspond to individuals who are \textbf{S}usceptible, \textbf{E}xposed but not infectious, \textbf{A}symptomatic and infectious, \textbf{I}nfectious and symptomatic, or \textbf{R}ecovered. The \bi\ compartment is further subdivided into individuals who are not yet symptomatic (i.e. presymptomatic) and those who are fully symptomatic, denoted \bia\ and \bib\ respectively. See \citet{Mar15} for a detailed overview of compartment models for disease.

\subsection{Infection Dynamics}
\label{sec:inf_dyn}

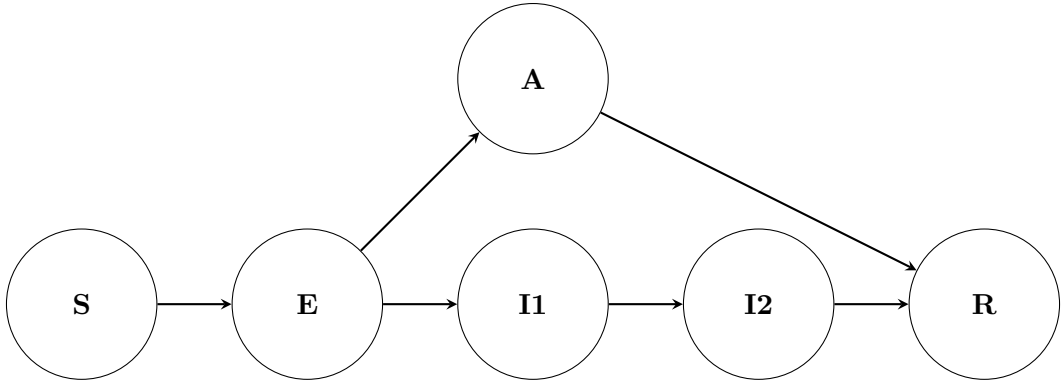
\begin{figure}[tp]
	\centering
\begin{tikzpicture}[node distance = 3cm]
	\node (S) [compartment] {\bs};
	\node (E) [compartment, right of=S] {\be};
	\node (I1) [compartment, right of=E] {\bia};
	\node (I2) [compartment, right of=I1] {\bib};
	\node (A) [compartment, above of=I1] {\ba};
	\node (R) [compartment, right of=I2] {\br};
	\draw [arrow] (S) -- (E);
	\draw [arrow] (E) -- (A);
	\draw [arrow] (E) -- (I1);
	\draw [arrow] (I1) -- (I2);
	\draw [arrow] (A) -- (R);
	\draw [arrow] (I2) -- (R);
\end{tikzpicture}
\caption{Modelled disease trajectory. Arrows represent possible transitions.}
\label{fig:Trans}
\end{figure}

Figure \ref{fig:Trans} shows which transitions are allowed in our model. In short, susceptible individuals can only transition to exposed. Exposed individuals transition to either asymptomatic or presymptomatic. The asymptomatic individuals transition directly to recovered, whereas those who are presymptomatic will transition through symptomatic before finally becoming recovered.

We model holding times in the \be, \ba, \bia\ and \bib\ compartments using geometric random variables (supported on the positive integers, excluding zero), with a different success probability for each compartment. Call these probabilities $q_E$, $q_A$, $q_{I1}$ and $q_{I2}$ respectively. Thus, the number of individuals transitioning out of compartment $X$ on any particular day follows the Binomial$(N_X, q_X)$ distribution, where $X$ is a compartment other than \bs\ or \br\ (see below for details on transitioning out of \bs; no transitions out of \br\ occur), and $N_X$ is the number of individuals in compartment $X$ on that day. The specific individuals who transition out of a compartment are chosen uniformly at random from the members of that compartment. 

We also use Bernoulli random variables to choose a destination when individuals transition out of \be. Call $q_{EA}$ the probability that a transition from \be\ is to \ba. Thus, among those individuals transitioning out of \be, the number that transition to \ba\ follows a binomial distribution. The remainder of those leaving \be\ enter compartment \bia.

Holding times in the \bs\ compartment are more complicated. Specifically, the probability of a susceptible individual transitioning to exposed on a particular day depends on both the sizes of classes in which the susceptible is enrolled and the number of contagious individuals who are also enrolled in these classes. 

Consider a class with one susceptible student, and some number of students in the contagious compartments, \ba, \bia\ and \bib. We assume that all possible transmission events are independent. On a particular day, each contagious individual has some probability of transmitting the disease to our susceptible student. This probability depends on which compartment the contagious individial is in. We model the transmission probability in a class between a single susceptible-contagious pair as inversely proportional to the square root of the class size, with a different proportionality constant for each contagious compartment: $\theta_A$, $\theta_{I1}$ and $\theta_{I2}$. Letting $\tau_X$ be the pairwise transmission probability for a contagious individual in compartment $X$, the infection probability for a single susceptible on a particular day is then $\tau_* = 1 - (1 - \tau_A)^{M_A} (1 - \tau_{I1})^{M_{I1}} (1 - \tau_{I2})^{M_{I2}}$, where $M_X$ is the number of individuals in the class who are in compartment $X$. Finally, the number of new cases in this class follows the Binomial$(M_S, \tau_*)$ distribution. After determining how many individuals will transition out of \bs\ on a particular day, we select the particular individuals uniformly at random from this compartment. 

The process just described covers how to generate transitions in a single class. To simulate a full day, we run this process independently in every class that meets on that day. Note that it is possible under our framework for an individual to become infected in more than one of their classes. When this happens, we simply move this individual to compartment \be, and ignore any multiplicity effects.

The model described above has eight parameters, four probabilities for geometric holding time distributions, one probability for Bernoulli trials to choose transition destinations, and three proportionality constants for transmission probabilities. To simplify identification of parameter values from the literature, we re-parameterize the infectiousness parameters for compartments \ba\ and \bia\ to be proportional to the infectiousness of individuals in compartment \bib. That is, we write $\theta_A = \rho_A \theta_{I2}$ or, equivalently, $\tau_A = \rho_A \tau_{I2}$. We define $\rho_{I1}$ similarly. 

We also consider a control strategy where classes above a certain size, $\phi$, are moved online, thereby preventing transmission between students in these classes\footnotemark. We arbitrarily choose the threshold values 20, 50, 100 and $\infty$, as these are qualitatively different class sizes (the maximum class size is 481). After removing classes above the specified threshold from the network, we find the largest component of the new network and remove any students who are not connected to this main component. Table \ref{tab:network_sizes} gives the number of students remaining in both the full network and the largest connected component after removing classes above the specified threshold. We focus on only the largest component of each network; the size of the full network is included only for completeness. The threshold size is another parameter for our model, giving a total of nine.

\footnotetext{Recall that we treat tutorials and labs independently of the courses with which they are associated. We also apply this strategy to the removal of classes from the network. That is, it is possible for the lecture portion of a course to be moved online while labs continue to meet in person. This is consistent with SFU's early strategy of prioritizing in-person metting of classes with experiential components \citep{SFU20III,SFU21}}

\begin{table}[tp]
	\centering
	\begin{tabular}{c|c|c}
		Threshold & Size of Network & Size of Largest Component\\
		\hline
		$20$ & 17,851 & 16,866\\
		$50$ & 25,470 & 23,660\\
		$100$ & 26,540 & 24,752\\
		$\infty$ & 27,307 & 25,627
	\end{tabular}
	\caption{Sizes of networks for various class size thresholds, both before and after removing isolated components.}
	\label{tab:network_sizes}
\end{table}

Table \ref{tab:par_vals} lists candidate values and their sources for each of our parameters. For each parameter related to disease progression, we use three plausible values based on a literature review. See Appendix \ref{app:par_vals} for more details.

\begin{table}[tp]
	\centering
	\begin{tabular}{c|c|c}
		Parameter & Values & Source\\
		\hline
		$\theta_{I2}$ & 0.141, 0.198, 0.240 & \citet{Tho21}\\
		$\rho_A$ & 0.4, 0.75, 1 & \citet{Joh21}\\
		$\rho_{I1}$ & 0.18, 0.63, 2.26 & \citet{Bui20}\\
		$q_E$ & 0.168, 0.182, 0.196 & \citet{Xin21}\\
		$q_A$ & 0.115, 0.138, 0.169 & \citet{Byr20}\\
		$q_{I1}$ & 0.333, 0.435, 0.833 & \citet{Bya20, Xin21}\\
		$q_{I2}$ & 0.063, 0.075, 0.092 & \citet{Bya20} \\
		$q_{EA}$ & 0.09, 0.18, 0.26 & \citet{Bya20}\\
		$\phi$ & 20, 50, 100, $\infty$ &
	\end{tabular}
	\caption{Candidate values and their sources for each parameter in our model.}
	\label{tab:par_vals}
\end{table}

\subsection{Simulation Details}

Our simulation is run in discrete time over a period of \numDays\ days, which corresponds roughly to a semester at SFU. On each day, we simulate the dynamics described in Section \ref{sec:inf_dyn}. On a particular day, new cases can only arise in classes which meet on that day, but we do allow all individuals who are infected to (possibly) progress to a more advanced stage of the disease. We initialize our simulation by randomly moving \numInitialCases\ individuals to the \bib\ compartment\footnotemark (i.e. sympomatic infected). At each time step, we track the number of individuals in each compartment.

\footnotetext{We use \numInitialCases\ instead of 1 initial case beause we want to investigate properties of outbreaks. A similar study with 1 initial case would be better able to investigate the probability of an outbreak occurring, at the expense of having less data for studying the outbreaks themselves. We discuss this in more detail in Section \ref{sec:extensions}.}

We repeat our simulation \numReps\ times at each parameter combination. This gives us a total of \numParCombs\ sets of \numReps\ disease trajectories. 

\section{Analysis}
\label{sec:analysis}

In this section, we describe the analysis we perform on the output of our simulation. This includes the selection of a small number of summary statistics of the disease trajectories, as well as the associated descriptive analysis and modelling of these summaries. The results of our analysis are presented in Section \ref{sec:results}, and interpretation is presented in Section \ref{sec:discussion}.

\subsection{Summarizing the Trajectories}

To avoid characterizing the entire trajectories simultaneously, we summarize each curve with a pair of statistics: the proportion of the population who ever becomes infected, and the peak infection size. The former, defined as the proportion of individuals who leave compartment \bs\ by the end of term, is referred to in the epidemiology literature as the cumulative incidence of infection, or CII \citep{Cow20}. The latter measure, peak infection size, is defined as the largest proportion of individuals simultaneously outside compartments \bs\ and \br\ (i.e. the proportion of individuals among \be, \ba, \bia\ and \bib). While the peak infection size is closely connected to CII, the CII measures impact of the disease across the entire term, while peak infection size measures the largest instantaneous number of cases. 

Both of our summaries are defined as proportions of the population size. However, there is some ambiguity in the definition of these proportions, since the number of students changes for different class size thresholds. We also restrict attention to the largest connected component in each network, so the population size is not even the number of students remaining after thresholding. Unless stated otherwise, when we discuss a proportion or a population size, it is taken with respect to the number of students in the largest connected component after thresholding.

\subsection{Statistical Analysis}
\label{sec:stat_anal}

Recall that the purpose of our analysis is to provide interpretable results to help inform policy decisions. As such, our modelling choices favor ease of interpretation over statistical optimality. 

The analyses of our two response variables is similar, so we decribe the common methodology here. We begin by constructing side-by-side boxplots for each simulation parameter, summarizing the distribution of the response within each parameter level. This gives a preliminary qualitative understanding of the marginal relationships between simulation parameters and the response. 

For both summaries, the difference across levels of $\phi$, the class size threshold, is much greater than across levels of the other parameters. As such, we emphasize the effect of class size threshold as a predictor throughout our analysis. Since the differences are so large across thresholds, we produce a histrogram of the outcome at each threshold level.

One extreme outlier was detected for both outcomes in the threshold = 100 group. This outbreak has an order of magnitude fewer infections that the other simulation runs with the same parameter settings. This behaviour is due to the infection being slow to get going (although the outbreak never becomes extinct). Because of its wildly different behaviour from similar simulation runs, we opt to remove the outlier from analysis. We do, however, retain the other runs at this parameter combination.

Next, we fit a regression tree model to explain the mean response using our parameters as covariates (see, e.g., \citealp{Bri84}). Briefly, a regression tree recursively partitions the predictor space by choosing a predictor and a value of that predictor with which to divide the space into `low' and `high'. Within each new subregion, the response is predicted by its mean over the sample observations in that region. The choice of predictor and dividing value is made to minimize the global sum of squared errors over all such splits. This partitioning process is then applied repeatedly, with splits at each step being chosen from among those that could be made in any of the subregions defined up to that point (i.e.\ recursive partitioning). The result of the recursive partitioning algorithm can be visualized in a tree shape, starting with a `root' node, and with each split replacing an existing node with two `child nodes'. It is common to continue splitting until a very large tree is produced (which almost certainly overfits the data), then to choose a smaller subtree using cross validation\footnotemark. The selection of a subtree is evocatively called `pruning'.

\footnotetext{\label{foot:CV}For an introduction to cross validation and more details, see \citet{Has09}. When choosing subtrees, we first find the tree with minimum average error across CV folds. Call this the CV-min tree. Then, we compute the standard deviation of the CV error across folds at the minimizer, and select the smallest tree with mean CV error no larger than minimum plus 1 standard error. Call this the CV-1se tree.}

Since both of our responses are proportions, before doing any splitting we apply a logit transformation (i.e.\ $x \mapsto \log[x / (1-x)]$) to get a range more compatible with the squared error loss used by regression trees. Fitting is done using the \texttt{rpart} package in \texttt{R} \citep{The19}. As mentioned above, class size threshold is much more strongly associated with CII than any epidemiological parameters, so we divide our data into groups based on the class size threshold and fit a separate model within each group.

At each threshold level, we begin by fitting a large tree, then prune this tree using the usual cross-validation (CV) strategy\footnote{See footnote \ref{foot:CV}}. We also consider pruning to each of a small number of sizes which lend themselves well to interpretation. Specifically, we consider trees with 10, 25, 50, 100 and 200 splits. To illustrate the behaviour of these small trees, we plot the trajectory of root mean squared error (CV-RMSE) across tree sizes up to 200. The performance of these trees gets reasonably close to that of the minimum CV-RMSE tree.

Next, we report measures of variable importance and goodness-of-fit for the interpretable and CV-optimal trees discussed above. See \citet{The19} and \citet{Bri84} for details of how variable importance is measured with regression trees. For goodness-of-fit, we report the CV-RMSE of each tree. Note that, since the tuned trees are chosen to optimize the CV-RMSE, this performance metric is biased for those trees and should be interpreted cautiously (see, e.g., \citealp{Has09}).

Finally, for each class size threshold we choose a small tree which performs fairly well, and explore which splits are actually made. Since this conveys similar information to the variable importances discussed above but with more detail, we include plots of the splits made for the chosen trees in Appendix \ref{app:Splits}.

\section{Results}
\label{sec:results}

In this section, we present the output of our analysis without any discussion. See Section \ref{sec:discussion} for interpretation of these results. We first present all results for the cumulative incidence of infection (CII), then move on to peak outbreak size.

\subsection{Cumulative Incidence of Infection}

\begin{figure}[tp]
	\centering
	\includegraphics[width=\textwidth]{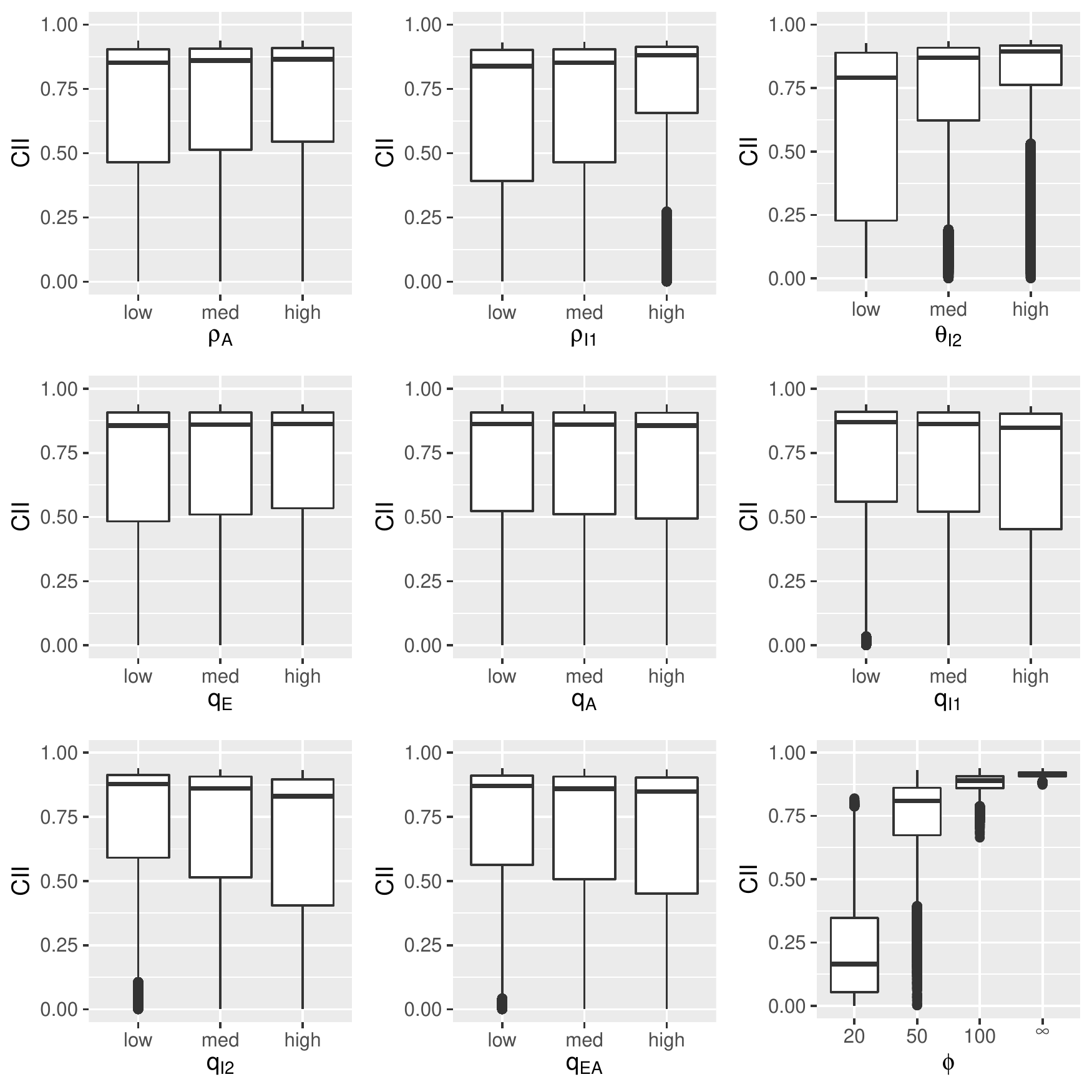}
	\caption{Boxplots of CII across levels for each simulation parameter.}
	\label{fig:CII_box}
\end{figure}

In Figure \ref{fig:CII_box}, we give boxplots of the CII across levels for each of the simulation parameters. 

\begin{figure}[tp]
	\centering
	\includegraphics[width=\textwidth]{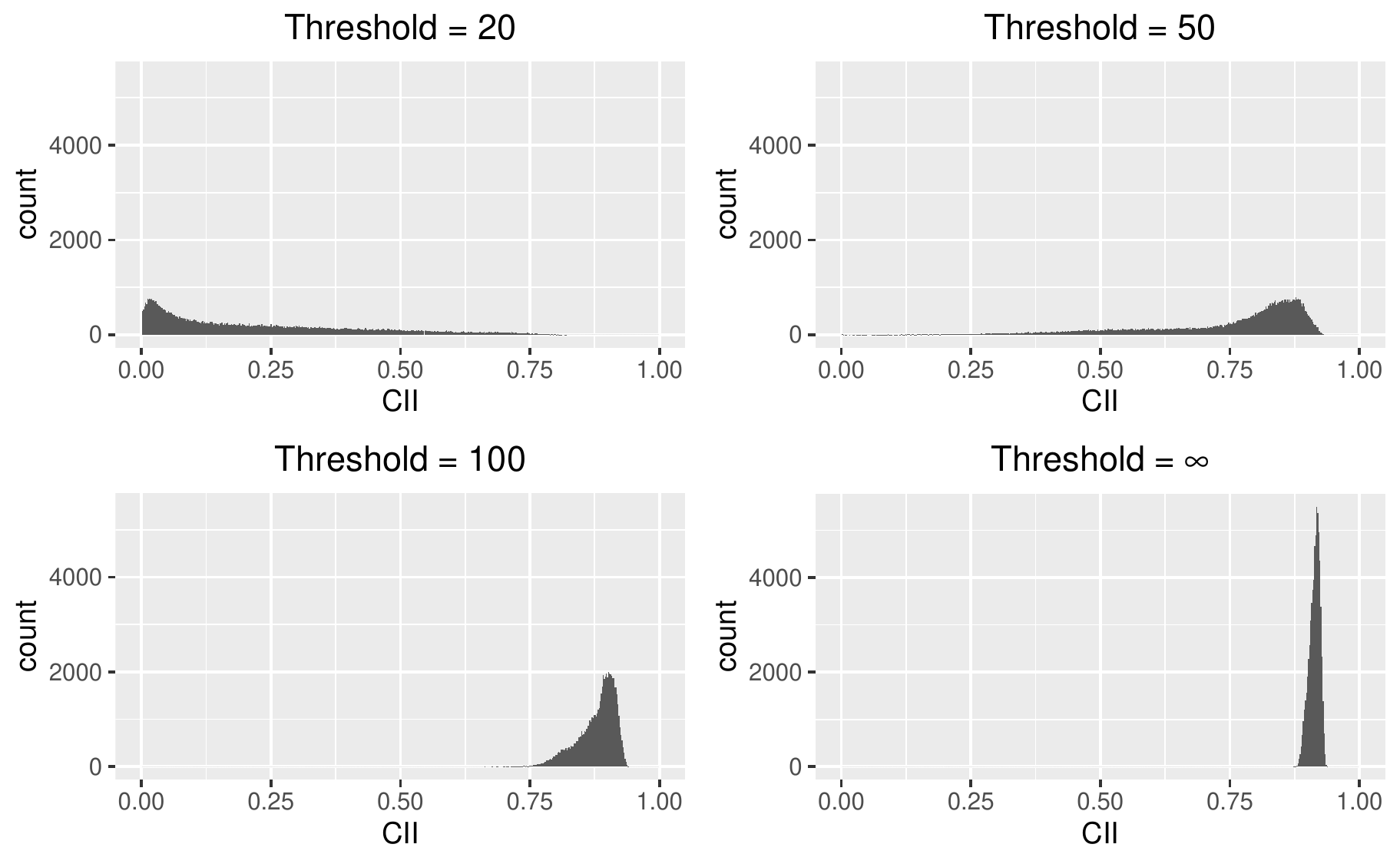}
	\caption{Histograms of CII within each class size threshold. Axis scales held fixed across plots.}
	\label{fig:CII_hist_homo}
\end{figure}

\begin{figure}[tp]
	\centering
	\includegraphics[width=\textwidth]{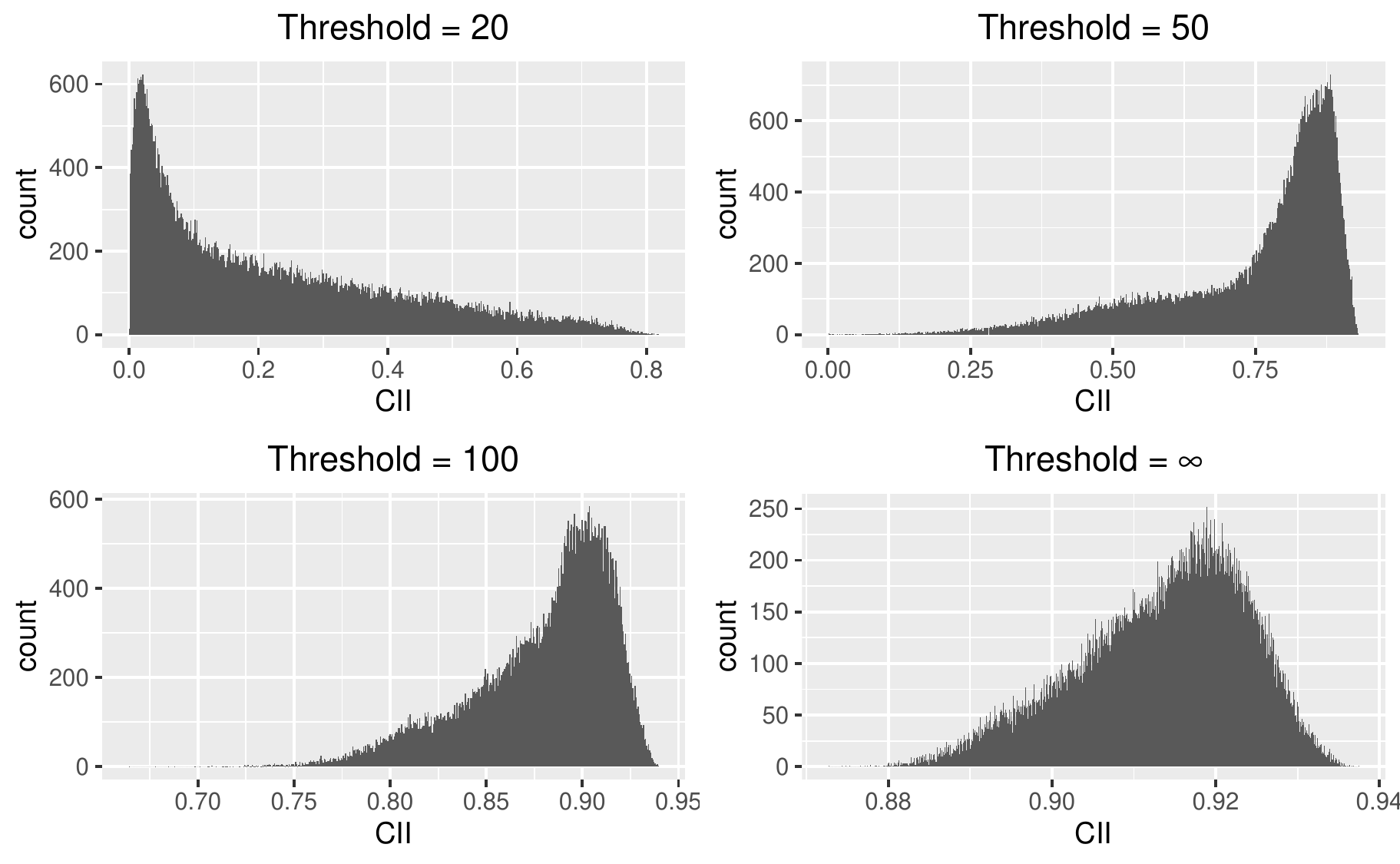}
	\caption{Histograms of CII within each class size threshold. Axis scales differ across plots.}
	\label{fig:CII_hist_hetero}
\end{figure}

Figures \ref{fig:CII_hist_homo} and \ref{fig:CII_hist_hetero} give histograms of the CII for each class size threshold. Axis scales are held fixed in Figure \ref{fig:CII_hist_homo} and allowed to vary between plots in Figure \ref{fig:CII_hist_hetero}. These two histograms highlight respectively the differences between threshold levels and features of the distribution within each threshold.

Table \ref{tab:CII_CV_Trees} gives the number of splits and the root mean squared CV error (CV-RMSE) for both CV trees across all class size thresholds when predicting logit-CII. Recall that the logit transformation is applied to the response before any model fitting is performed, and that CV-RMSEs are reported on the logit-scale.

\begin{table}[tp]
	\centering
	\begin{tabular}{c|rr|rr}
	  \hline
		& \multicolumn{2}{c|}{CV-1se} & \multicolumn{2}{c}{CV-min} \\
	 Threshold & Splits & CV-RMSE & splits & CV-RMSE \\ 
	  \hline
	  20 & 184 & 0.64 & 348 & 0.64 \\ 
	  50 & 591 & 0.22 & 5660 & 0.21 \\ 
	  100 & 1751 & 0.05 & 3629 & 0.05 \\ 
	  $\infty$ & 520 & 0.04 & 854 & 0.04 \\
	   \hline
	\end{tabular}
	\caption{Summaries of CV-tuned trees for predicting logit-CII across class size thresholds.} 
	\label{tab:CII_CV_Trees}
	\end{table}

\begin{figure}[tp]
	\centering
	\begin{tabular}{cc}
		\includegraphics[width=0.4\textwidth]{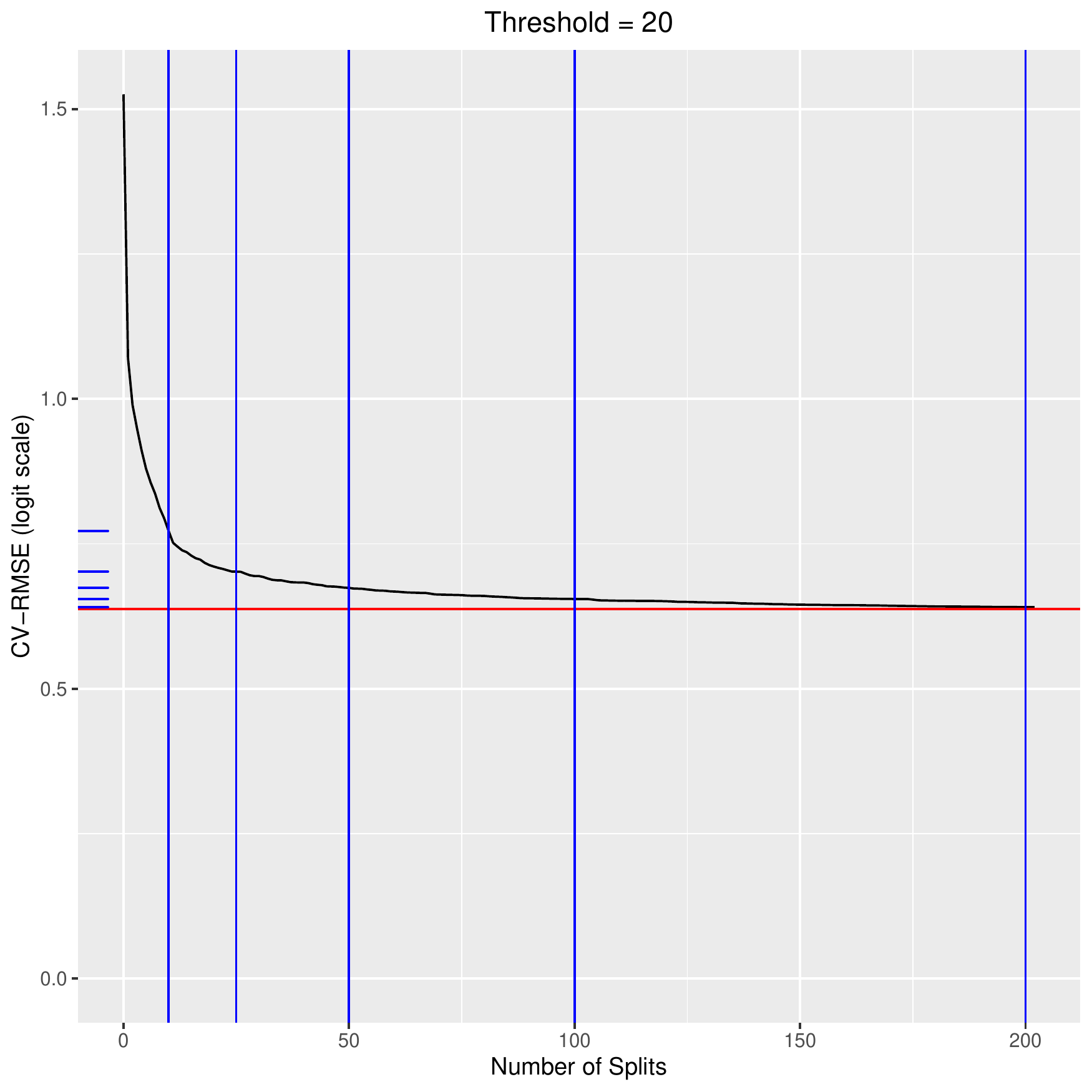} & \includegraphics[width=0.4\textwidth]{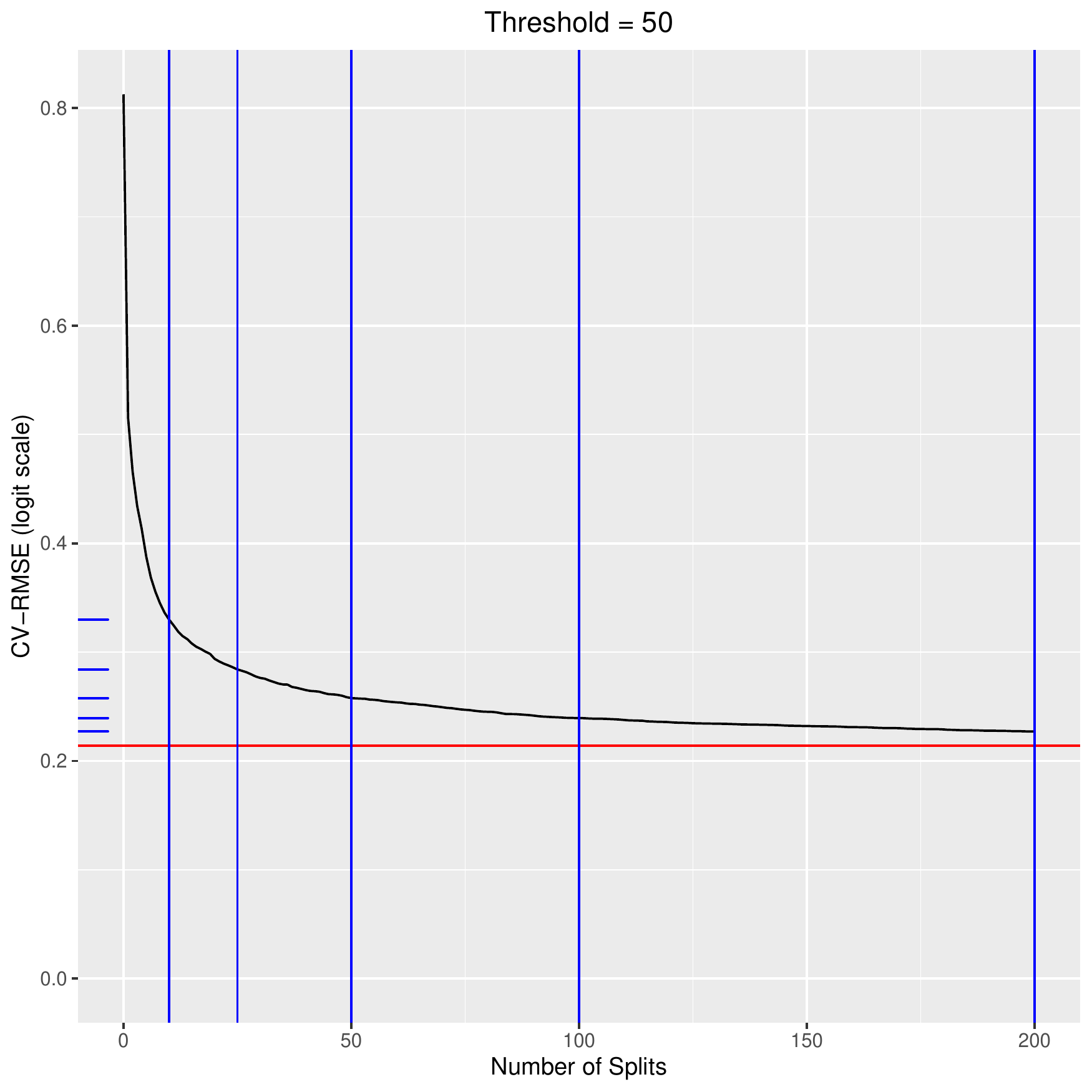}\\
		\includegraphics[width=0.4\textwidth]{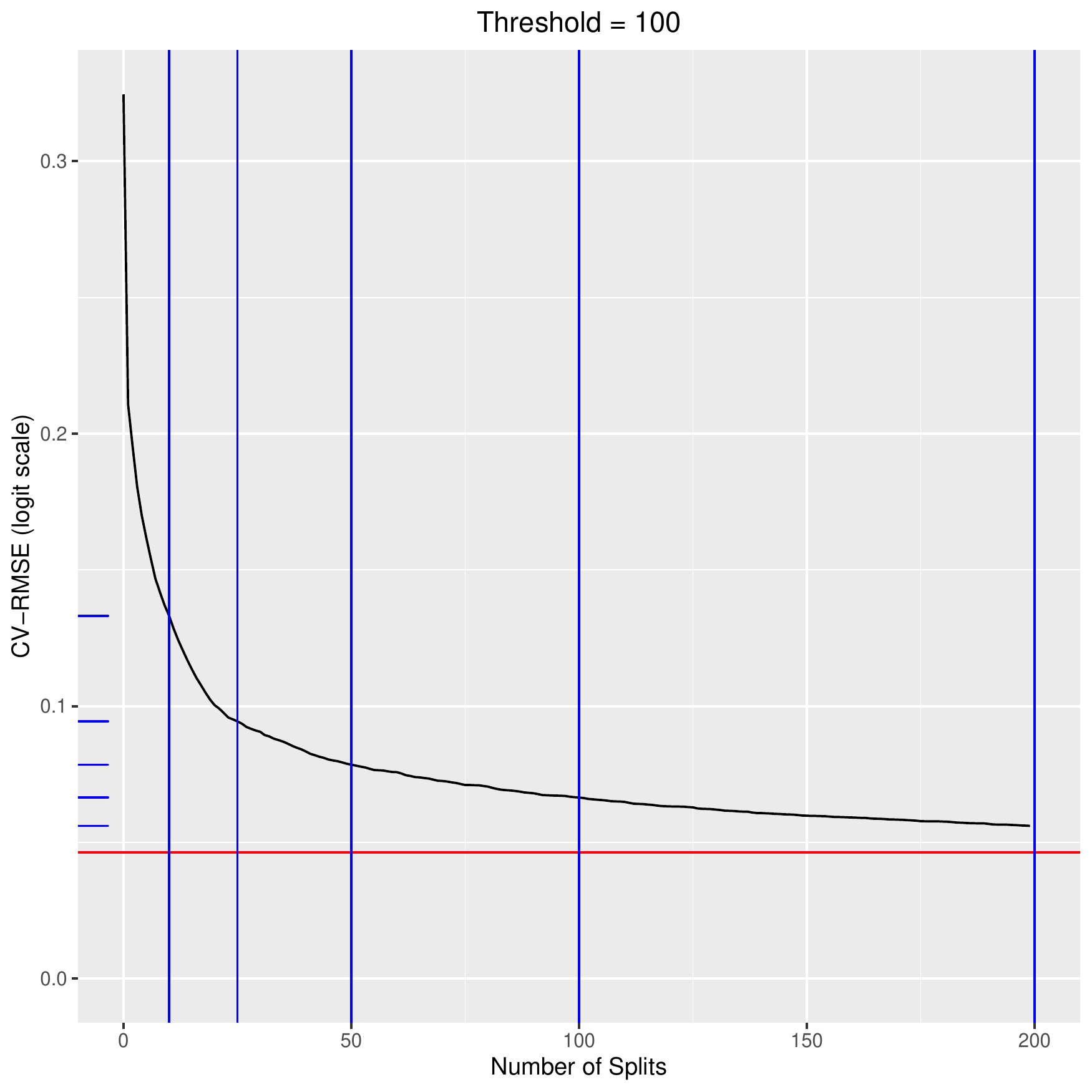} & \includegraphics[width=0.4\textwidth]{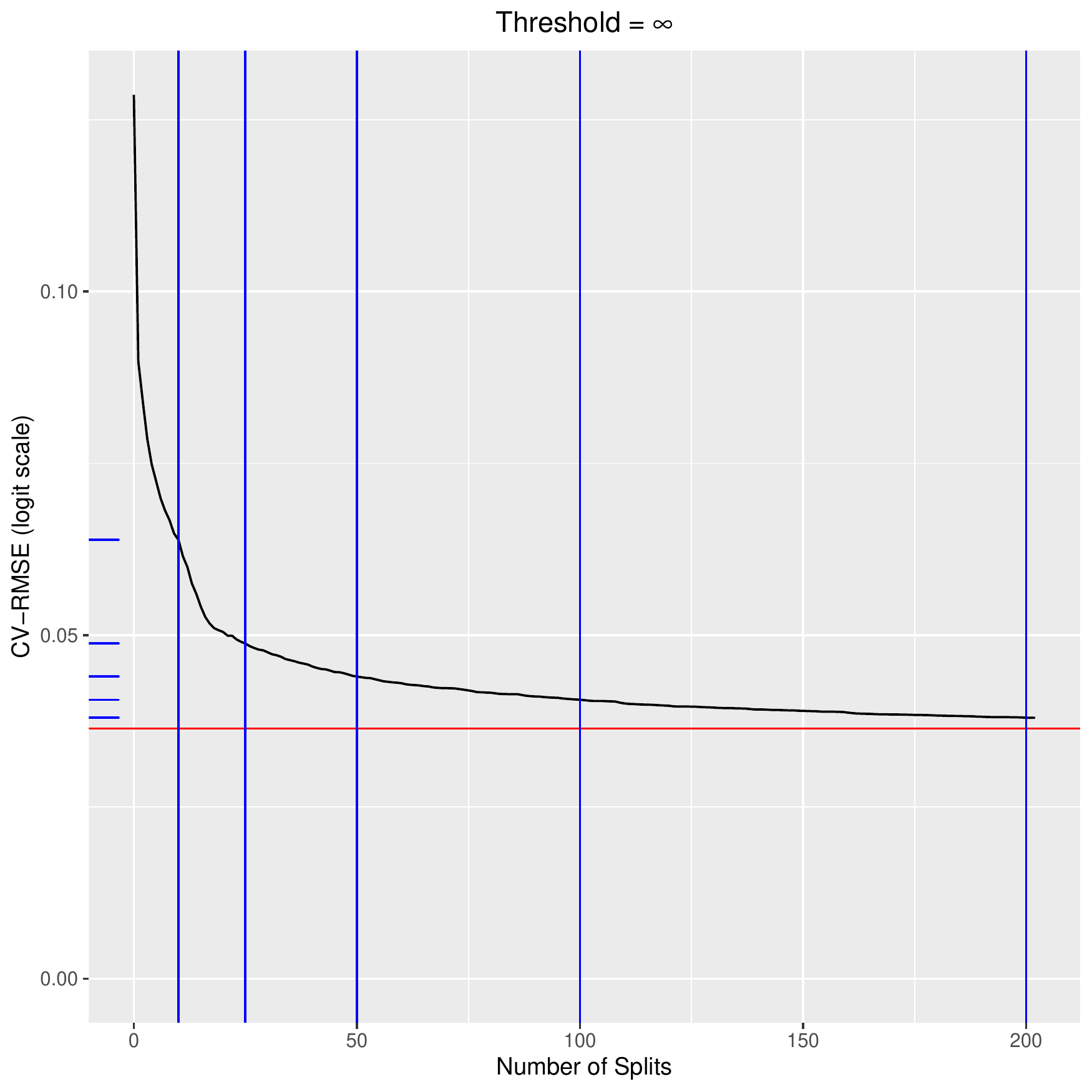}
	\end{tabular}
	\caption{CV-RMSE for predicting logit-CII across tree sizes for each class size threshold. Vertical lines correspond to trees with 10, 25, 50, 100 and 200 splits, with ticks on the Y-axis at these trees' CV-RMSE values. The horizontal line is the global minimum.}
	\label{fig:CII_err_focused}
\end{figure}

Figure \ref{fig:CII_err_focused} gives the logit-CII CV-RMSE as a function of tree size, focusing attention on trees with few splits (i.e. at most 200). Vertical lines are given at 10, 25, 50, 100 and 200 splits, and ticks on the Y-axis show these trees' error rates. The global CV-RMSE is given by a horizontal line.

Table \ref{tab:CII_var_imp} gives variable importance measures for some trees at each class size threshold when predicting logit-CII. The values for each tree have been re-scaled to sum to one across variables. See Section \ref{sec:inf_dyn} for parameter definitions.

\begin{table}[tp]
	\centering
	\begin{tabular}{c|c|cccccccc}
		\hline
 		Threshold & Tree & $\rho_A$ & $\rho_{I1}$ & $\theta_{I2}$ & $q_E$ & $q_A$ & $q_{I1}$ & $q_{I2}$ & $q_{EA}$ \\ 
  	\hline
		\multirow{7}{*}{20} & 10 &  & 0.12 & 0.78 &  &  & 0.06 & 0.04 &  \\ 
		&  25 &  & 0.11 & 0.73 &  &  & 0.06 & 0.09 & $\approx 0$ \\ 
		&  50 & $\approx 0$ & 0.11 & 0.72 &  &  & 0.06 & 0.09 & 0.01 \\ 
		&  100 & 0.01 & 0.11 & 0.71 &  & $\approx 0$ & 0.06 & 0.09 & 0.02 \\ 
		&  200 & 0.01 & 0.11 & 0.70 & $\approx 0$ & $\approx 0$ & 0.06 & 0.09 & 0.02 \\ 
		&  CV-1se & 0.01 & 0.11 & 0.70 & $\approx 0$ & $\approx 0$ & 0.06 & 0.09 & 0.02 \\ 
		&  CV-min & 0.01 & 0.11 & 0.69 & $\approx 0$ & $\approx 0$ & 0.06 & 0.09 & 0.02 \\ 
	\hline
		\multirow{7}{*}{50} & 10 &  & 0.12 & 0.77 &  &  & 0.05 & 0.07 &  \\ 
		& 25 & $\approx 0$ & 0.13 & 0.73 &  &  & 0.05 & 0.09 & 0.01 \\ 
		& 50 & 0.01 & 0.13 & 0.71 &  &  & 0.05 & 0.09 & 0.02 \\ 
		& 100 & 0.01 & 0.13 & 0.70 &  &  & 0.05 & 0.09 & 0.02 \\ 
		& 200 & 0.01 & 0.13 & 0.69 & $\approx 0$ & $\approx 0$ & 0.05 & 0.09 & 0.03 \\ 
		& CV-1se & 0.02 & 0.13 & 0.68 & $\approx 0$ & $\approx 0$ & 0.05 & 0.08 & 0.03 \\ 
		& CV-min & 0.02 & 0.13 & 0.68 & 0.01 & 0.01 & 0.06 & 0.08 & 0.03 \\ 
	\hline
		\multirow{7}{*}{100} & 10 &  & 0.07 & 0.76 &  &  & 0.03 & 0.14 &  \\ 
		& 25 & $\approx 0$ & 0.10 & 0.69 &  &  & 0.05 & 0.14 & 0.01 \\ 
		& 50 & 0.01 & 0.10 & 0.67 &  &  & 0.05 & 0.14 & 0.02 \\ 
		& 100 & 0.01 & 0.10 & 0.66 &  &  & 0.06 & 0.14 & 0.03 \\ 
		& 200 & 0.01 & 0.10 & 0.65 &  & $\approx 0$ & 0.06 & 0.14 & 0.03 \\ 
		& CV-1se & 0.02 & 0.10 & 0.65 & $\approx 0$ & $\approx 0$ & 0.06 & 0.14 & 0.03 \\ 
		& CV-min & 0.02 & 0.10 & 0.65 & $\approx 0$ & $\approx 0$ & 0.06 & 0.14 & 0.03 \\ 
	\hline
		\multirow{7}{*}{$\infty$} & 10 &  & 0.10 & 0.73 &  &  & 0.02 & 0.15 &  \\ 
  		& 25 &  & 0.11 & 0.67 &  &  & 0.06 & 0.15 & 0.01 \\ 
  		& 50 & 0.01 & 0.10 & 0.65 &  &  & 0.06 & 0.16 & 0.02 \\ 
  		& 100 & 0.01 & 0.11 & 0.63 &  &  & 0.07 & 0.15 & 0.03 \\ 
  		& 200 & 0.01 & 0.11 & 0.62 &  & $\approx 0$ & 0.07 & 0.15 & 0.03 \\ 
  		& CV-1se & 0.02 & 0.11 & 0.62 & $\approx 0$ & $\approx 0$ & 0.07 & 0.15 & 0.04 \\ 
  		& CV-min & 0.02 & 0.11 & 0.62 & $\approx 0$ & $\approx 0$ & 0.07 & 0.15 & 0.04 \\ 
   	\hline	
	\end{tabular}
	\caption{Variable importance measures for selected trees of interest in each class size threshold for predicting logit-CII. Values of $\approx 0$ round to 0. Blank cells indicate that no splits were made on that variable by that tree.}
	\label{tab:CII_var_imp}
\end{table}

Table \ref{tab:CII_GOF} gives the CV-RMSEs of some selected trees at each class size threshold when predicting logit-CII. As discussed in Section \ref{sec:stat_anal}, reported CV-RMSEs for the CV-1se and CV-min trees are biased due to the optimization involved in selecting these trees.

\begin{table}[tp]
\centering
\begin{tabular}{c|ccccccc}
  \hline
 Threshold & 10 & 25 & 50 & 100 & 200 & CV-1se* & CV-min* \\ 
  \hline
20 & 0.77 & 0.70 & 0.67 & 0.66 & 0.64 & 0.64 & 0.64 \\ 
  50 & 0.33 & 0.28 & 0.26 & 0.24 & 0.23 & 0.22 & 0.21 \\ 
  100 & 0.13 & 0.09 & 0.08 & 0.07 & 0.06 & 0.05 & 0.05 \\ 
  $\infty$ & 0.06 & 0.05 & 0.04 & 0.04 & 0.04 & 0.04 & 0.04 \\ 
   \hline
\end{tabular}
\caption{CV-RMSE for predicting logit-CII using selected trees across class size thresholds. *CV-RMSEs for trees chosen based on this metric are optimistically biased.} 
\label{tab:CII_GOF}
\end{table}

Taken together, the above results suggest that 25 splits provides a good balance between interpretability and capturing most of the possible improvement in CV-RMSE when predicting logit-CII. Ideally, we would use a larger tree, but adding more splits quickly makes the tree infeasible to visualize and interpret. Futhermore all trees CV-RMSE values are quite small. See Appendix \ref{app:Splits} for the pruned trees with 25 splits at each threshold level.

All trees split first on $\theta_{I2}$, the infectiousness parameter for symptomatic cases. For threshold levels of 20 and 50, the next step includes splitting up whichever pair of $\theta_{I2}$ levels remained together after the first split. These trees then split on $\rho_{I1}$, the relative infectiousness of presymptomatic individuals, followed by the holding time parameters for presymptomatic and sympomatic individuals. For threshold levels of 100 and $\infty$, the second stage splits on $q_{I2}$, the holding time parameter for symptomatic individuals. These trees then split on $\theta_{I2}$ if possible (i.e. in the group with two remaining levels of this predictor), and on $\rho_{I1}$. There is remarkable similarity between the trees fit at threshold levels of 20 and 50, as well as between thresholds of 100 and $\infty$.

\subsection{Peak Outbreak Size}

\begin{figure}[tp]
	\centering
	\includegraphics[width=\textwidth]{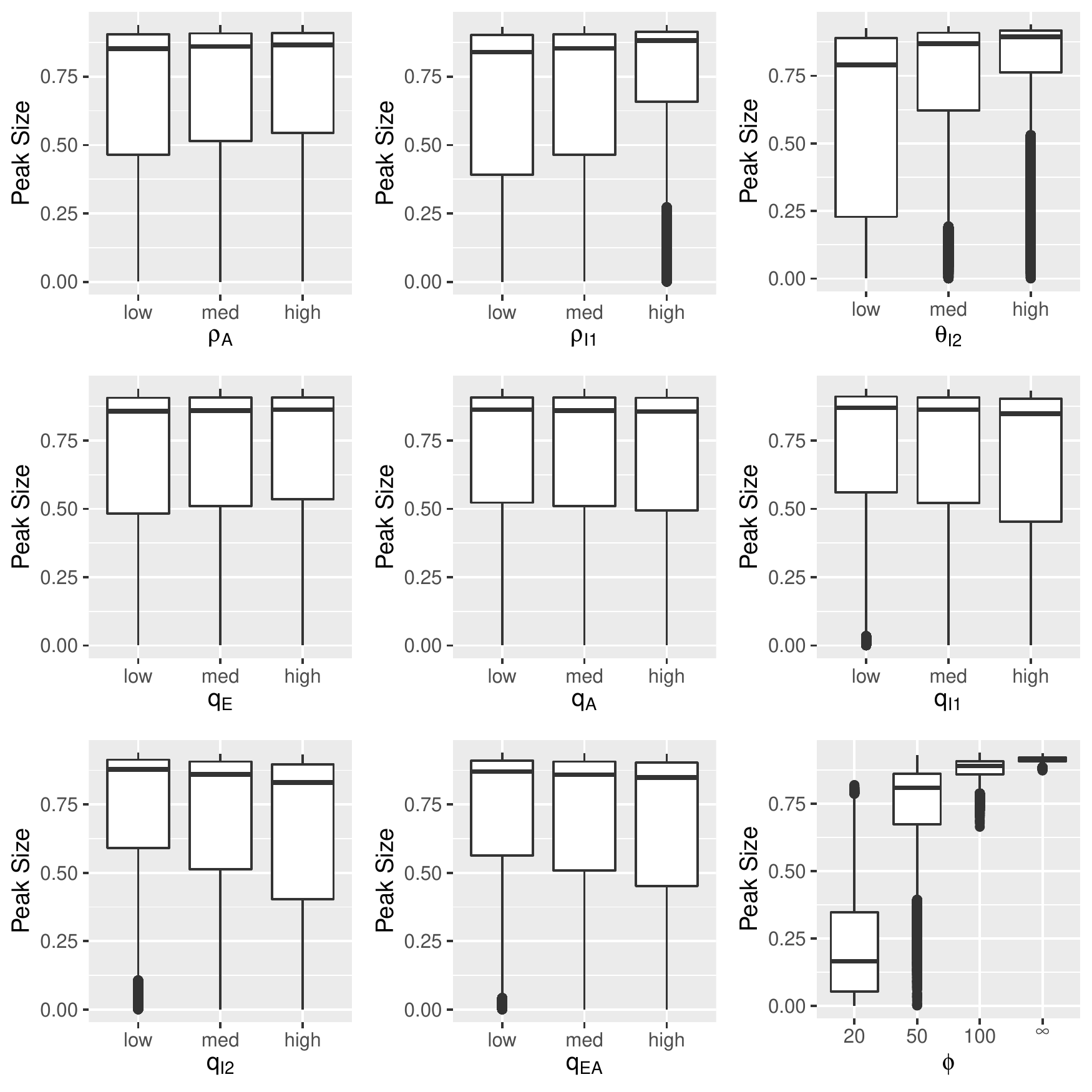}
	\caption{Boxplots of peak outbreak size across levels for each simulation parameter.}
	\label{fig:peak_box}
\end{figure}

In Figure \ref{fig:peak_box}, we give boxplots of the peak outbreak size across levels for each of the simulation parameters. 

\begin{figure}[tp]
	\centering
	\includegraphics[width=\textwidth]{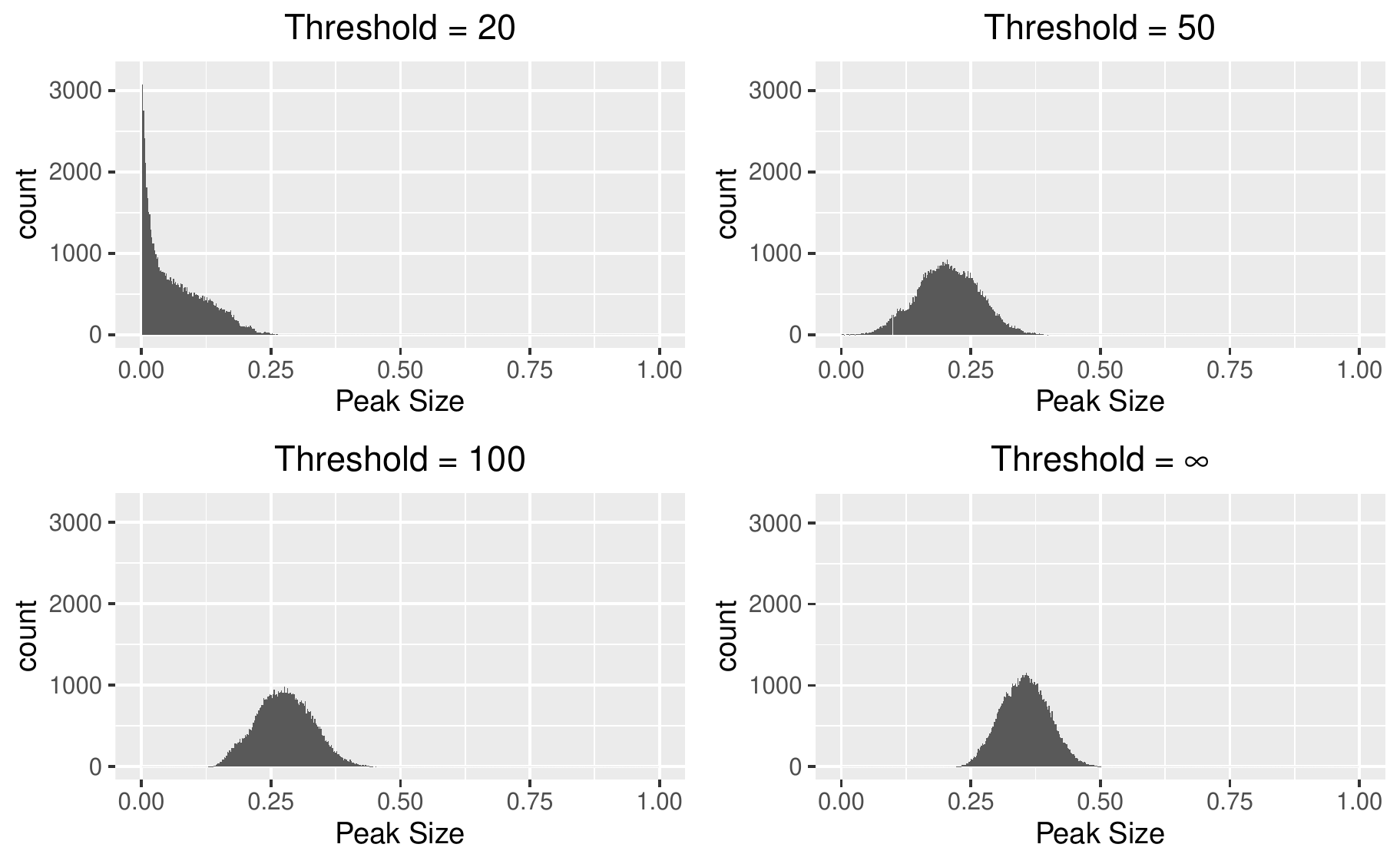}
	\caption{Histograms of peak outbreak size within each class size threshold. Axis scales held fixed across plots.}
	\label{fig:peak_hist_homo}
\end{figure}

\begin{figure}[tp]
	\centering
	\includegraphics[width=\textwidth]{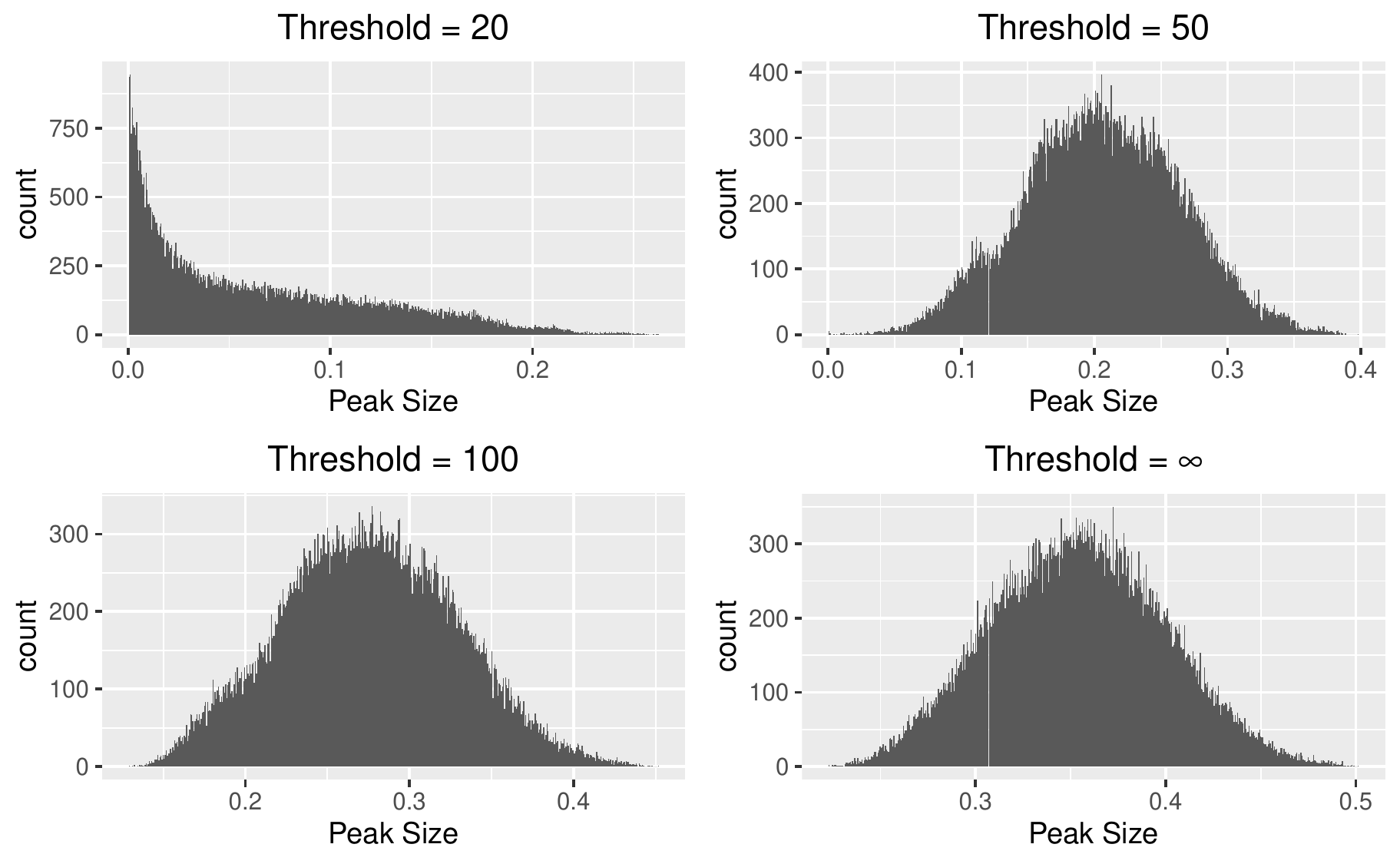}
	\caption{Histograms of peak outbreak size within each class size threshold. Axis scales differ across plots.}
	\label{fig:peak_hist_hetero}
\end{figure}

Figures \ref{fig:peak_hist_homo} and \ref{fig:peak_hist_hetero} give histograms of the peak outbreak size for each class size threshold. Axis scales are held fixed in Figure \ref{fig:peak_hist_homo} and allowed to vary between plots in Figure \ref{fig:peak_hist_hetero}. These two histograms highlight respectively the differences between threshold levels and features of the distribution within each threshold.

Table \ref{tab:peak_CV_Trees} gives the number of splits and the CV-RMSE for both CV trees across all class size thresholds when predicting logit-peak outbreak size.

\begin{table}[tp]
	\centering
	\begin{tabular}{c|rr|rr}
	  \hline
		& \multicolumn{2}{c|}{CV-1se} & \multicolumn{2}{c}{CV-min} \\
	 Threshold & Splits & CV-RMSE & splits & CV-RMSE \\ 
	  	\hline
		20 & 223 & 0.59 & 508 & 0.59 \\ 
  		50 & 565 & 0.11 & 6228 & 0.10 \\ 
  		100 & 4616 & 0.03 & 6057 & 0.03 \\ 
  		$\infty$ & 4485 & 0.02 & 6033 & 0.02 \\
	   \hline
	\end{tabular}
	\caption{Summaries of CV-tuned trees for predicting logit-peak outbreak size across class size thresholds.} 
	\label{tab:peak_CV_Trees}
	\end{table}

\begin{figure}[tp]
	\centering
	\begin{tabular}{cc}
		\includegraphics[width=0.4\textwidth]{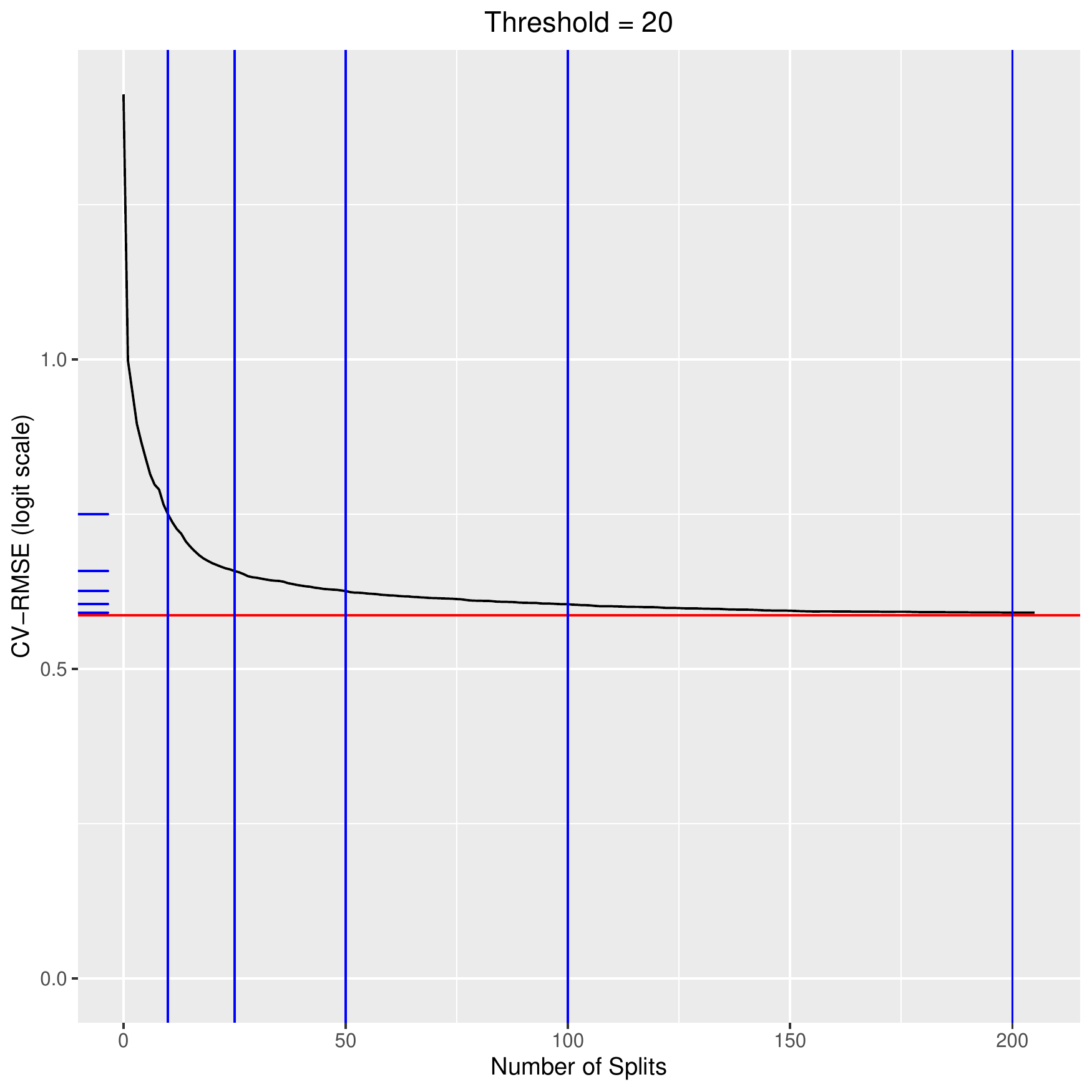} & \includegraphics[width=0.4\textwidth]{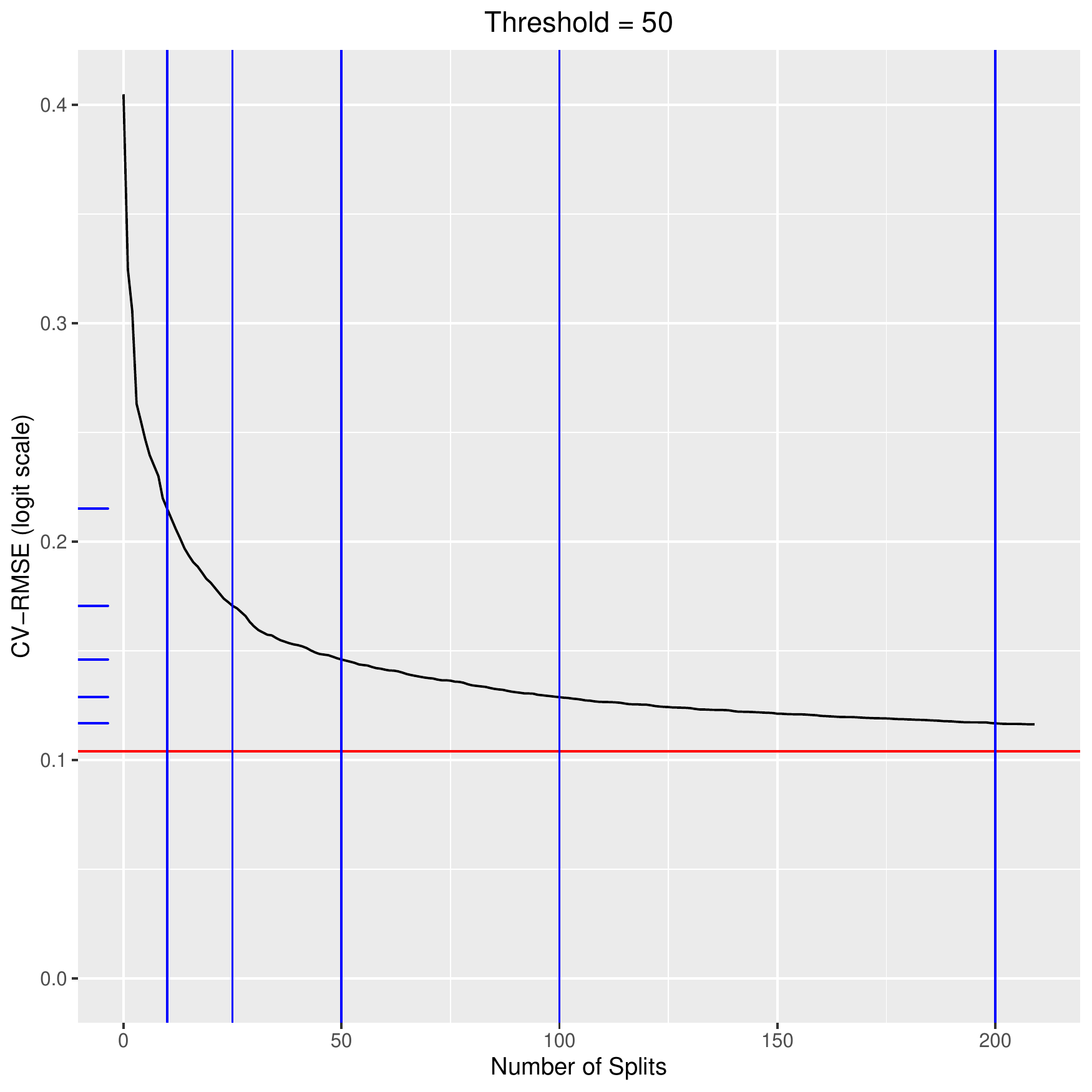}\\
		\includegraphics[width=0.4\textwidth]{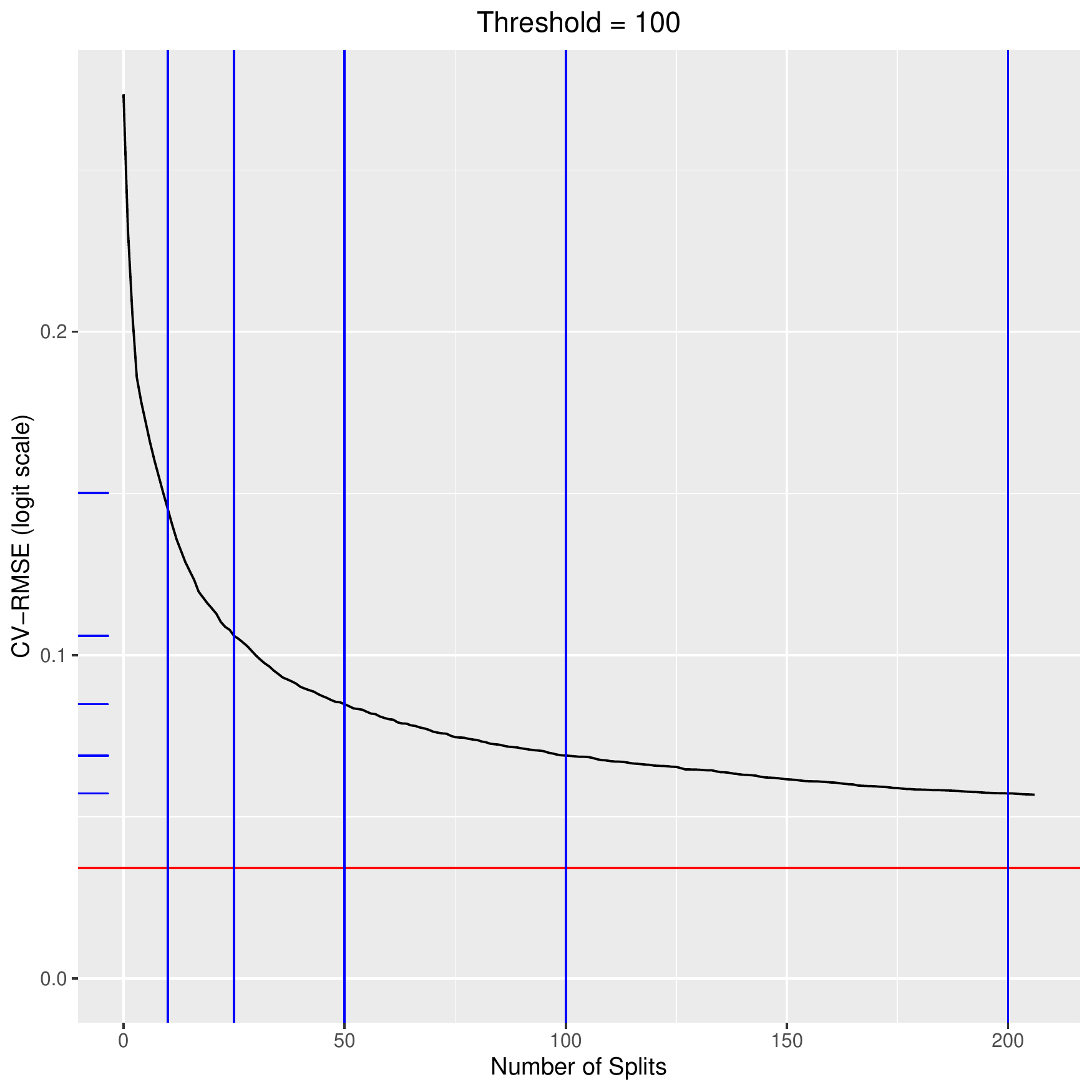} & \includegraphics[width=0.4\textwidth]{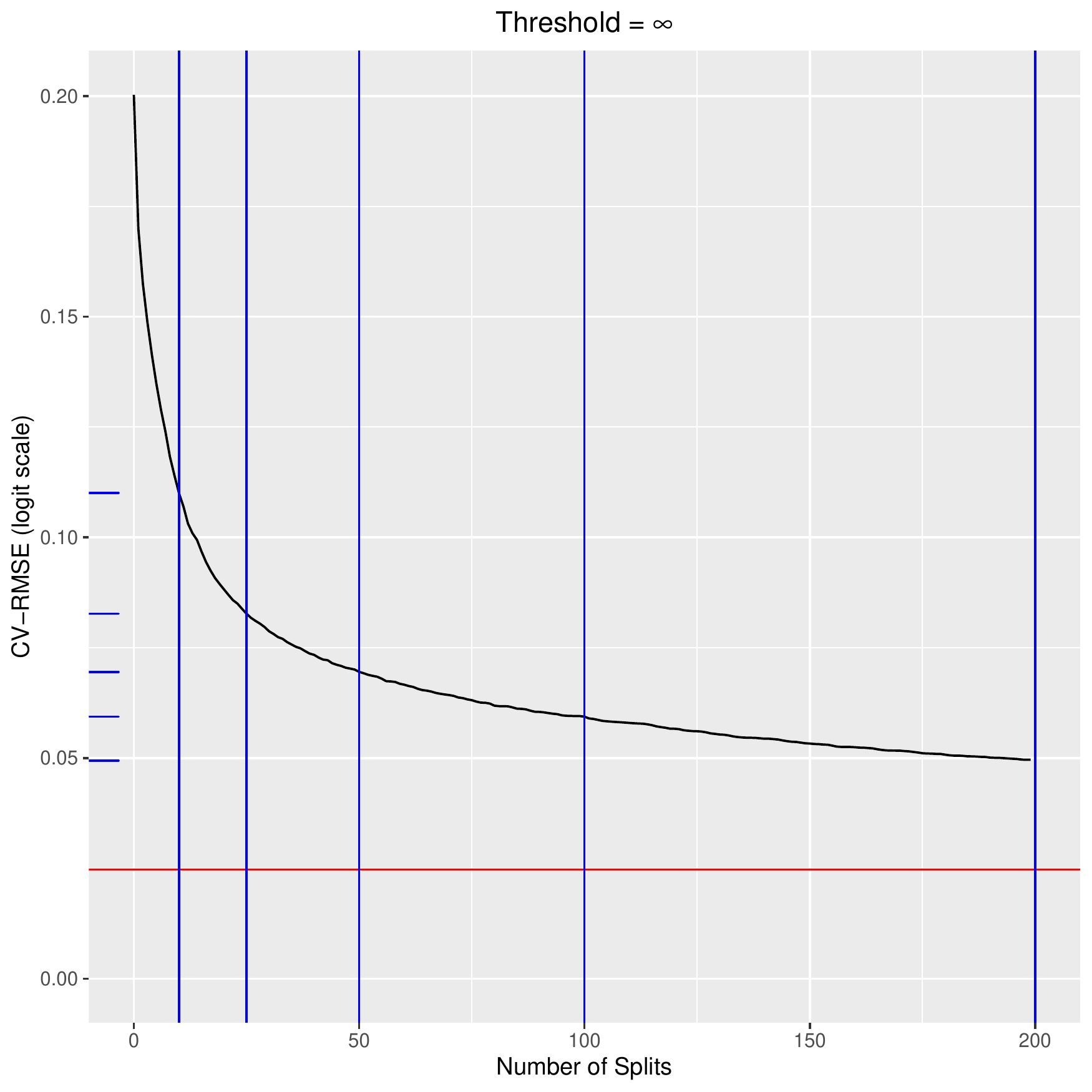}
	\end{tabular}
	\caption[CV-RMSE for predicting logit-peak outbreak size across tree sizes for each class size threshold.]{CV-RMSE for predicting logit-peak outbreak size across tree sizes for each class size threshold. Vertical lines correspond to trees with 10, 25, 50, 100 and 200 splits\footnotemark, with ticks on the Y-axis at these trees' CV-RMSE values. The horizontal line is the global minimum.}
	\label{fig:peak_err_focused}
\end{figure}

Figure \ref{fig:peak_err_focused} gives the logit-peak outbreak size CV-RMSE as a function of tree size, focusing attention on trees with few splits (i.e. at most 200). Vertical lines are given at 10, 25, 50, 100 and 200 splits\footnotemark, and ticks on the Y-axis show these trees' error rates. The global CV-RMSE is given by a horizontal line.

\footnotetext{\label{foot:splits}All trees with 10 splits are worse than the best tree with 9 splits with respect to the criterion used for tuning. As such, the optimal 9-split tree is used in place of a 10-split tree. For consistency, we still refer to this as the 10-split tree.}

\footnotetext{See footnote \ref{foot:splits}.}

Table \ref{tab:peak_var_imp} gives variable importance measures for some trees at each class size threshold when predicting logit-peak outbreak size. The values for each tree have been re-scaled to sum to one across variables. See Section \ref{sec:inf_dyn} for parameter definitions.

\begin{table}[tp]
	\centering
	\begin{tabular}{c|c|cccccccc}
		\hline
 		Threshold & Tree & $\rho_A$ & $\rho_{I1}$ & $\theta_{I2}$ & $q_E$ & $q_A$ & $q_{I1}$ & $q_{I2}$ & $q_{EA}$ \\ 
  	\hline
		\multirow{7}{*}{20} & 10 &  & 0.08 & 0.77 &  &  & 0.02 & 0.13 &  \\ 
		& 25 &  & 0.09 & 0.71 &  &  & 0.06 & 0.14 & 0.01 \\ 
		& 50 & $\approx 0$ & 0.09 & 0.69 &  &  & 0.06 & 0.14 & 0.02 \\ 
		& 100 & 0.01 & 0.09 & 0.68 &  & $\approx 0$ & 0.06 & 0.14 & 0.03 \\ 
		& 200 & 0.01 & 0.09 & 0.67 & $\approx 0$ & $\approx 0$ & 0.06 & 0.14 & 0.03 \\ 
		& CV-1se & 0.01 & 0.09 & 0.67 & $\approx 0$ & $\approx 0$ & 0.06 & 0.14 & 0.03 \\ 
		& CV-min & 0.01 & 0.09 & 0.66 & $\approx 0$ & $\approx 0$ & 0.06 & 0.14 & 0.03 \\ 
	\hline
		\multirow{7}{*}{50} & 10 &  & 0.08 & 0.50 &  &  & 0.05 & 0.37 &  \\ 
		& 25 &  & 0.10 & 0.46 &  &  & 0.08 & 0.33 & 0.03 \\ 
		& 50 & 0.01 & 0.10 & 0.44 &  &  & 0.09 & 0.31 & 0.05 \\ 
		& 100 & 0.01 & 0.11 & 0.43 &  & $\approx 0$ & 0.10 & 0.30 & 0.06 \\ 
		& 200 & 0.01 & 0.11 & 0.42 & $\approx 0$ & $\approx 0$ & 0.10 & 0.29 & 0.07 \\ 
		& CV-1se & 0.01 & 0.11 & 0.41 & $\approx 0$ & 0.01 & 0.10 & 0.29 & 0.07 \\ 
		& CV-min & 0.02 & 0.10 & 0.41 & 0.01 & 0.01 & 0.10 & 0.28 & 0.06 \\ 
	\hline
		\multirow{7}{*}{100} & 10 &  & 0.07 & 0.40 &  &  & 0.07 & 0.45 &  \\ 
		& 25 &  & 0.13 & 0.34 &  &  & 0.12 & 0.37 & 0.04 \\ 
		& 50 &  & 0.12 & 0.34 &  &  & 0.12 & 0.35 & 0.06 \\ 
		& 100 & 0.00 & 0.12 & 0.33 &  & 0.00 & 0.13 & 0.34 & 0.08 \\ 
		& 200 & 0.00 & 0.12 & 0.33 & 0.00 & 0.01 & 0.13 & 0.33 & 0.08 \\ 
		& CV-1se & 0.01 & 0.12 & 0.32 & 0.01 & 0.02 & 0.12 & 0.32 & 0.08 \\ 
		& CV-min & 0.01 & 0.12 & 0.32 & 0.01 & 0.02 & 0.12 & 0.32 & 0.08 \\ 
	\hline
		\multirow{7}{*}{$\infty$} & 10 &  & 0.16 & 0.22 &  &  & 0.22 & 0.40 &  \\ 
		& 25 &  & 0.14 & 0.20 &  &  & 0.18 & 0.41 & 0.07 \\ 
		& 50 &  & 0.13 & 0.19 &  & $\approx 0$ & 0.18 & 0.40 & 0.10 \\ 
		& 100 &  & 0.13 & 0.19 & $\approx 0$ & 0.01 & 0.18 & 0.38 & 0.10 \\ 
		& 200 & $\approx 0$ & 0.13 & 0.19 & 0.02 & 0.01 & 0.18 & 0.37 & 0.10 \\ 
		& CV-1se & 0.01 & 0.12 & 0.19 & 0.03 & 0.02 & 0.17 & 0.35 & 0.10 \\ 
		& CV-min & 0.01 & 0.12 & 0.19 & 0.03 & 0.02 & 0.17 & 0.35 & 0.10 \\
   	\hline	
	\end{tabular}
	\caption{Variable importance measures for selected trees of interest in each class size threshold for predicting logit-peak outbreak size. Values of $\approx 0$ round to 0. Blank cells indicate that no splits were made on that variable by that tree.}
	\label{tab:peak_var_imp}
\end{table}

Table \ref{tab:peak_GOF} gives the CV-RMSEs of some selected trees at each class size threshold when predicting logit-peak outbreak size. As discussed in Section \ref{sec:stat_anal}, reported CV-RMSEs for the CV-1se and CV-min trees are biased due to the optimization involved in selecting these trees.

\begin{table}[tp]
\centering
\begin{tabular}{c|ccccccc}
  \hline
 Threshold & 10 & 25 & 50 & 100 & 200 & CV-1se* & CV-min* \\ 
  \hline
  20 & 0.75 & 0.66 & 0.63 & 0.61 & 0.59 & 0.59 & 0.59 \\ 
  50 & 0.21 & 0.17 & 0.15 & 0.13 & 0.12 & 0.11 & 0.10 \\ 
  100 & 0.15 & 0.11 & 0.08 & 0.07 & 0.06 & 0.03 & 0.03 \\ 
  $\infty$ & 0.11 & 0.08 & 0.07 & 0.06 & 0.05 & 0.02 & 0.02 \\ 
   \hline
\end{tabular}
\caption{CV-RMSE for predicting logit-peak outbreak size using selected trees across class size thresholds. *CV-RMSEs for trees chosen based on this metric are optimistically biased.} 
\label{tab:peak_GOF}
\end{table}

The above results suggest that 25 splits provides a good balance between interpretability and capturing most of the possible improvement in CV-RMSE when predicting logit-peak outbreak size. Although the relative difference in CV-RMSE between the 25 split and CV-min trees appears large in Figure \ref{fig:peak_err_focused}, the absolute difference is quite small (see, e.g. Table \ref{tab:peak_GOF} and the corresponding results for CII in Table \ref{tab:CII_GOF}). Ideally, we would use a larger tree, but adding more splits quickly makes the tree infeasible to visualize and interpret. Futhermore all trees CV-RMSE values are quite small. See Appendix \ref{app:Splits} for the pruned trees with 25 splits at each threshold level.

When predicting peak outbreak size, the three thresholded groups (i.e. threshold level of 20, 50 or 100) are quite similar, while the unthresholded group is different. In all threshold levels other than $\infty$, the first split is on $\theta_{I2}$, the infectiousness parameter for symptomatic individuals. However, in the unthresholded group, the first split is on $q_{I2}$, the holding time parameter for symptomatic individuals. At the next step, splits are made either on $q_{I2}$ for the thresholded trees, or on $\theta_{I2}$ and $q_{I1}$, the holding time parameter for presymptomatic individuals, when no thresholding is applied. The trees start to diverge at the next level, with splits being made on $\theta_{I2}$, $q_{I1}$, $q_{I2}$ and $\rho_{I1}$, the relative infectiousness of presymptomatic to symptomatic individuals.

\section{Discussion}
\label{sec:discussion}

The results of our simulation show that moving classes online is strongly associated with lessening the severity of a disease outbreak. Conversely, the differences across levels of any single epidemiological parameter are small when averaged across the other parameters. This is true whether severity is measured by total number infected (a.k.a.\ cumulative incidence of infection, or CII) or by peak simultaneous case count. Differences across class size thresholds is most pronounced for the CII, see Figures \ref{fig:CII_box}-\ref{fig:CII_hist_hetero}, where if all classes are allowed to meet in person, most students become infected over the course of a term. However, if all classes with more than 20 students are moved online, most simulation runs have well below $50\%$ infections. The effect is weaker for peak infection size, see Figures \ref{fig:peak_box}-\ref{fig:peak_hist_hetero}, but it is still clearly preferable to keep large classes online from this perspective.

To elaborate, for both CII and peak outbreak size, we see a strong qualitative difference between a threshold of 20 versus the other levels. Specifically, thresholding at 20 gives a right-tailed distribution with most of the mass concentrated near 0. Increasing the threshold gives either a left-tailed or symmetric distribution for CII or peak outbreak size respectively, with values concentrated away from 0. Unsurprisingly, as the threshold level increases, the distribution moves farther from 0. Said differently, with more classes allowed to meet in person, more students become infected, regardless of whether total or peak case numbers are being counted. A cursory analysis (not shown) indicates that the low proportions of cases among more severe thresholding levels seen in Figure \ref{fig:CII_hist_homo} are because the outbreak did not have time to finish, not because it stalled (i.e. there are still many contagious individuals).

In our fitted tree models, most of the predictive power is captured by a small number of splits, relative to the performance of a full-sized tree (with the possible exception of the peak outbreak size at larger threshold levels, although the absolute differences there are small). This is fortunate, since one of the major advantages of regression trees is their interpretability, and this advantage is lost when one must try to interpret thousands of splits.

Relative contributions of the various epidemiological parameters differ across thresholds and across responses (see Tables \ref{tab:CII_var_imp} and \ref{tab:peak_var_imp}). For CII, we see most of the importance concentrated on one predictor; specifically, $p_{I2}$, the infectiousness parameter for symptomatic cases. For peak outbreak size, we start with most of the importance concentrated on $p_{I2}$ when thresholding at 20 students, but as we allow larger classes, $p_{I2}$ becomes less important and other variables become more important. Specifically, $q_{I2}$, the holding time parameter for the sympomatic compartment, matches the importance of $p_{I2}$ when thresholding at 100 students, and exceeds the importance of $p_{I2}$ when no thresholding is applied. We also see $\rho_{I1}$ and $q_{I1}$, the relative infectiousness and holding time parameters respectively for the presymptomatic compartment, become much more important relative to $\theta_{I2}$ as thresholding is weakened. These results are consistent with the actual splits made early in tree fitting; see Appendix \ref{app:Splits}.

A high importance score for $\theta_{I2}$ does not necessarily tell us that symptomatic cases are the primary driver of infection in our model. In fact, we expect $\theta_{I2}$ to appear important because we have parameterized the other compartments' infectiousness values relative to that of the symptomatic compartment. That is, if we change the infectiousness of sympomatic cases, we also change the infectiousness of the other contagious compartments, while the converse is not true. However, the high importance score of $q_{I2}$ for high threshold levels with peak outbreak size suggest that the symptomatic compartment is, in fact, an important determinant of the infection's peak severity over the course of a term.

Several variables either aren't selected for splitting or have a very low importance score: specifically, $\rho_A$, the relative infectiousness of asymptomatic cases, and $q_E$ and $q_A$, the holding time parameters for exposed status and asymptomatic cases respectively. The low importance of duration spent in the exposed compartment is unsurprising, since this compartment neither transmits nor receives infection. If exposed durations were often of a similar order to the duration of the simulation (\numDays\ days), then we would expect to see a larger effect, where many individuals never progress to the contagious phase of the disease. However, under our chosen parameter values, the mean time spent in the exposed compartment never exceeds 6 days. The low importance of parameters associated with asymptomatic cases is also unsurprising due to the relatively low infectiousness of individuals in this compartment (see Table \ref{tab:par_vals}). In fact, if we remove the single large value of $\rho_{I1}$, the relative infectiousness of presymptomatic cases, then the importance of this variable drops to be closer to that of $\rho_A$.

Our control strategy of removing classes above a specified threshold eliminates a substantial proportion of enrollments from the network. To ensure that any improvement seen is not only due to this `thinning' of paths along which the infection can spread, we repeat our simulation with enrollments removed uniformly at random instead of according to the more systematic class size threshold strategy. More specifically, for each class size threshold other than $\infty$, we remove enrollments chosen uniformly at random until the number remaining matches the number of enrollments among classes below that threshold. We call this process `thinning the network'. Note that thinning is applied after removing classes with only one student but before removing isolated components. Removing isolated components both before and after thinning leads to an excess of removed enrollments. Finally, we remove any isolated components and repeat the simulation as described in the rest of this section. The only difference is that here the parameter $\phi$ only takes values 20, 50 and 100, and is thought of as indexing the degree of thinning rather than as an explicit threshold (smaller values of $\phi$ correspond to smaller networks after thinning). Results of these simulations (not shown) suggest that our thresholding strategy is considerably more effective than thinning at random.

\subsection{Limitations}
\label{sec:limitations}
 Our study has some limitations which restrict the generalizability of its conclusions. Firstly is the source of the data. Our network is constructed using only data from a single university, SFU. Since different schools will have different enrollment networks, we do not necessarily expect our conclusions to generalize. However, the methodology we use is quite general, and other institutions could repeat our analysis to see whether similar conclusions hold there. Code used to perform our simulation and data analysis, as well as the enrollment data used to generate the network, are available in an accompanying GitHub repository \citep{Rut22}. 

 We now discuss some limitations of our dataset. This is not meant to be an exhaustive list, but rather to illustrate some of the challenges involved with modelling disease spread on a real population. To start, our network only links students through shared classes. As is clear from a cursory inspection of any university campus, classes are not the only way in which students interact. It is conceivable that we could incorporate data on living arrangements for students in residence, but no dataset could account for all the ways in which students meet for coffee, or stand near each other outside a classroom, or on a bus... In short, we cannot account for all the ways in which a disease can spread throughout the student population, so instead accept that we must limit our study (and therefore its conclusions) to the effect of transmission through shared courses.
 
 Another limitation is the implicit assumption that every student who is enrolled in a class attends every meeting of that class. This assumption is clearly not true. In fact, there may be systematic bias toward lower attendance for classes at less popular times (e.g. the earliest classes at SFU start at 8:30 am). SFU does not keep records of class attendance, so the data required to account for attendance in our model does not exist. Some work has been done to study rates of class attendance \citep{Dev96, vBl92}, but incorporating these models into our study is beyond the scope of this paper.
 
 The last limitation we discuss relates to class scheduling. At SFU, classes meet at the same times each week, typically for one or more hours on one or more days. We were only able to obtain data for the day(s) on which a class meets. This prevents us from accounting for the amount of time actually spent in a room with classmates. Given more detailed information, we could develop a model which more closely reflects real-world behaviour, but data privacy concerns limit the specificity of the data we are able to access.

\subsection{Extensions}
\label{sec:extensions}

In the previous section, we discussed some inherent limitations to our study based on the dataset we were provided. Here we briefly mention some ways our model could be expanded to incorporate other aspects of disease transmission, as well as ideas for related analysis which are of interest but beyond the scope of a single manuscript.

As was discussed in Section \ref{sec:limitations}, our model does not account for the possibility of infection outside of classes. While it would be impossible to fully model student behaviour, one might introduce a random number of infections at each time step. The addition of random infections from outside the disease model is referred to as a spark term and is discussed by \citet{Dea10}. These additional infections would represent out-of-class interactions that take place on-campus, as well as the possibility of contracting the disease somewhere off-campus. Random infections could be assigned uniformly across the susceptible population, or a separate model could be developed to describe students' heterogeneous risks of transmission outside classes.

Our simulation uses only three distinct values for each of the disease parameters due to the sharp increase in computational cost as more values are included. Future work might focus on a finer exploration of the parameter space.

There are many possible control measures to limit further spread by infectious individuals. Examples include mask wearing and not coming to class when sick. Masking can be incorporated into an existing model by reducing transmission rates. Other work suggests that mask use is important for reducing transmission risk \citep{Zho21}. One can also imagine numerous strategies for keeping sick students out of classes. Examples include quarantining individuals who feel sick, or moving individual classes online if any enrolled students show symptoms. While a more comprehensive control strategy which makes use of any of the methods described here or elsewhere (see, e.g., \citealp{Gre20}) will be more effective than any one measure in isolation, our work specifically illustrates the benefit to be gained by moving certain classes online.

Our statistical analysis is somewhat limited. While regression trees are interpretable, other methods often have better statistical properties \citep{Has09}. It would be interesting for future work to include a more detailed machine learning analysis of our simulation output focusing on prediction instead of interpretability. Such an analysis may uncover patterns that our tree model is unable to detect.

An important feature of stochastic disease modelling is whether a single infected individual produces a full outbreak, or whether the infected cases all recover before a critical number is reached (see, e.g., \citealp{Bri19}). This ``extinction probability'' is best studied empirically using a single intial infected individual, so our framework is ill-suited to measure this quantity (we always start with \numReps\ initial cases). One way this might be studied is to repeat our simulation with a single initial case and investigate the rate at which outbreaks go extinct before infecting many students.

\section{Conclusions}
\label{sec:conclusions}
It is clear from a cursory analysis of our simulation results that moving large classes online is an important tool in managing the risk of a large outbreak at SFU. More precisely, in order to ensure that large outbreaks are unlikely (e.g.\ those in which more than $50\%$ of students are infected over the course of the term), even moderately sized classes must be moved online. Applying a threshold value of 20 gives qualitatively different behaviour than any other level, with most students avoiding infection and the peak number of simultaneous cases being quite small (often below $1\%$ of students). From a policy perspective, this suggests that all but the smallest classes must be moved online in order to mitigate the chance of a severe outbreak.

We have also identified which disease parameters are most strongly associated with case counts. In particular, the total number of infections is driven mostly by the disease's transmissibility rather than by the duration of infectiousness. Conversely, the largest number of simultaneous cases is more heavily influenced by infection duration; particularly when more classes are allowed to meet in-person. This increased influence is especially pronounced for symptomatic cases, which is the stage of the disease when individuals are most infectious. These results suggest that efforts to reduce the duration of infectiousness, by quarantining for example, are best focused on symptomatic individuals. This is comforting since it is much more challenging to detect asymptomatic and presymptomatic cases.

\section*{Acknowledgements}

The authors would like to thank Mallory Gilmore for her thoughtful feedback on an early version of this manuscript, as well as to the staff at SFU who provided us with the data.

\section*{Declarations}

This work was funded by the Natural Science and Engineering Research Council of Canada. Data and code are available in an accompanying GitHub repository \citep{Rut22}. The authors have no known conflicts of interest with this work.

\begin{appendices}

\section{Parameter Values}
\label{app:par_vals}

Here, we describe how we chose values for our model parameters.

\subsection{Infectiousness Parameters}

\subsubsection{$\theta_{I1}$ - Proportionality constant for symptomatic spreaders}

\citet{Tho21} perform a meta-analysis of secondary attack rates across various settings, and between different sub-groups. Their pooled secondary attack rate for symptomatic individuals is 0.14, with $95\%$ CI of 0.10 to 0.17. Secondary attack rate is the probability of transmission between a particular pair. In our model, this is analogous to transmission in a class of two people, which has probability $\theta_I / \sqrt{2}$. Thus, we use values for $\theta_I$ of 0.141, 0.198 and 0.240.

\subsubsection{$\rho_A$ - Infectiousness of asymptomatic relative to symptomatic cases}

\citet{Joh21} give several estimates across multiple other studies for infectiousness of asymptomatic individuals relative to symptomatic ones. Taking a few of these, we get $\rho_A$ values of 0.4, 0.75 and 1.

\subsubsection{$\rho_{I2}$ - Infectiousness of presymptomatic relative to symptomatic cases}

\citet{Bui20} give an estimate for the relative risk of infection from presymptomatic individuals as 0.63, with a $95\%$ CI of 0.18 to 2.26. It's kind of hard to find. While the largest value here is quite extreme, there has been some discussion in the literature of peak transmissibility occurring before symptom onset (see, e.g., \citealp{He20}).

\subsection{Transition Parameters}

\subsubsection{$q_E$ - Pre-infectious}

\citet{Xin21} estimate the latent period to be 5.5 days, with a $95\%$ CI of $5.1-5.9$ days. Since the mean latent period in our model is $1/q_E$, we use values of  0.168, 0.182 and 0.196 for $q_E$.

\subsubsection{$q_A$ - Asymptomatic}

\citet{Byr20} give several estimates of the infectiousness period for asymptomatic cases. The most appropriate one for us is from \citet{Ma20}, which is 7.25 days, with a $95\%$ CI of $5.91 - 8.69$ days. This gives values for $q_A$ of 0.115, 0.138 and 0.169.

\subsubsection{$q_{I1}$ - Pre-sympomatic}

\citet{Xin21} also estimate the incubation period to be 6.9 days, with a $95\%$ CI of 6.3-7.5 days. Their estimated latent period for symptomatic cases is 5.5 days (5.1-6.0). Subtracting the latent period for symptomatic cases from the incubation period, we get estimated mean holding times in \bia\ of 1.2, 1.4 and 1.5 (subtracting bottom from bottom, middle from middle and top from top respectively). These in turn, give estimates for $q_{I1}$ of 0.667, 0.714 and 0.833.

Alternatively, \citet{Byr20} give several estimates for the duration of presymptomatic infectiousness. A nice one due to \citet{He20} is 2.3 days, with a $95\%$ CI of $0.8 - 3.0$ days. Our model does not allow for holding times of less than one day, so we replace the lower bound from \citet{Byr20} with the lower bound of 1.2 from \citet{Xin21}. 

Taken the above two studies together, we get values for $q_{I1}$ of 0.333, 0.435 and 0.833.

\subsubsection{$q_{I2}$ - Symptomatic}

\citet{Bya20} perform a meta-analysis of studies which report duration of symptomatic infectiousness. Many of these studies focus on hospitalized populations, since recovery is not easily defined (see their paper for more details). Their pooled estimate is 13.4 days, with a $95\%$ CI of 10.9 to 15.8 days. This gives values for $q_{I2}$ of 0.063, 0.075 and 0.092.

\subsection{$q_{EA}$ - Proportion of asymptomatic cases}

A meta-analysis by \citet{Bya20} estimate the proportion of asymptomatic cases to be $0.18$, with a $95\%$ CI of $0.09 - 0.26$. We use the estimate from their random effects model as it has a larger range of values. These authors also take care to ensure that studies included in their analysis correctly distinguish asymptomatic cases from presymptomatic ones.

\subsection{Summary}

See Table \ref{tab:appendix_par_vals} for a summary of our chosen parameter values.

\begin{table}[tp]
	\centering
	\begin{tabular}{c|c|c|c|c}
		Parameter & Min & Center & Max  & Source \\
		\hline
		$\theta_{I2}$ & 0.141 & 0.198 & 0.240 & \citet{Tho21}\\
		$\rho_A$ & 0.4 & 0.75 & 1 & \citet{Joh21}\\
		$\rho_{I1}$ & 0.18 & 0.63 & 2.26 & \citet{Bui20}\\
		$q_E$ & 0.168 & 0.182 & 0.196 & \citet{Xin21}\\
		$q_A$ & 0.115 & 0.138 & 0.169 & \citet{Byr20}\\
		$q_{I1}$ & 0.333 & 0.435 & 0.833 & \citet{Bya20, Xin21}\\
		$q_{I2}$ & 0.063 & 0.075 & 0.092 & \citet{Bya20} \\
		$q_{EA}$ & 0.09 & 0.18 & 0.26 & \citet{Bya20}
	\end{tabular}
	\caption{Chosen values for each parameter, along with relevant reference(s). See text for reasoning.}
	\label{tab:appendix_par_vals}
\end{table}

\section{Small Trees' Splits}
\label{app:Splits}

\begin{figure}[tp]
	\centering
	\includegraphics[width=\textwidth]{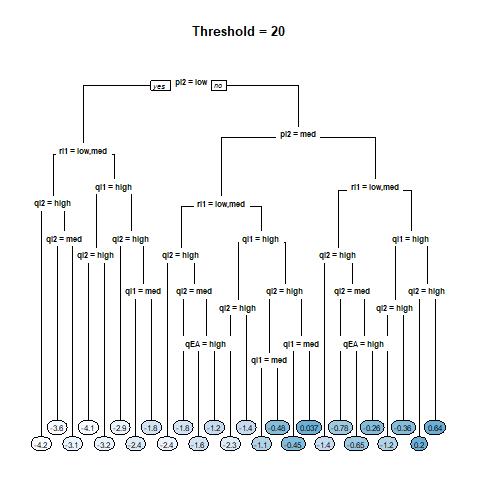}
	\caption{Pruned tree with 25 splits for predicting logit-CII with a class size threshold of 20. At each split, points which satisfy the listed condition move left, while the other points move right.}
	\label{fig:CII_splits_20}
\end{figure}

\begin{figure}[tp]
	\centering
	\includegraphics[width=\textwidth]{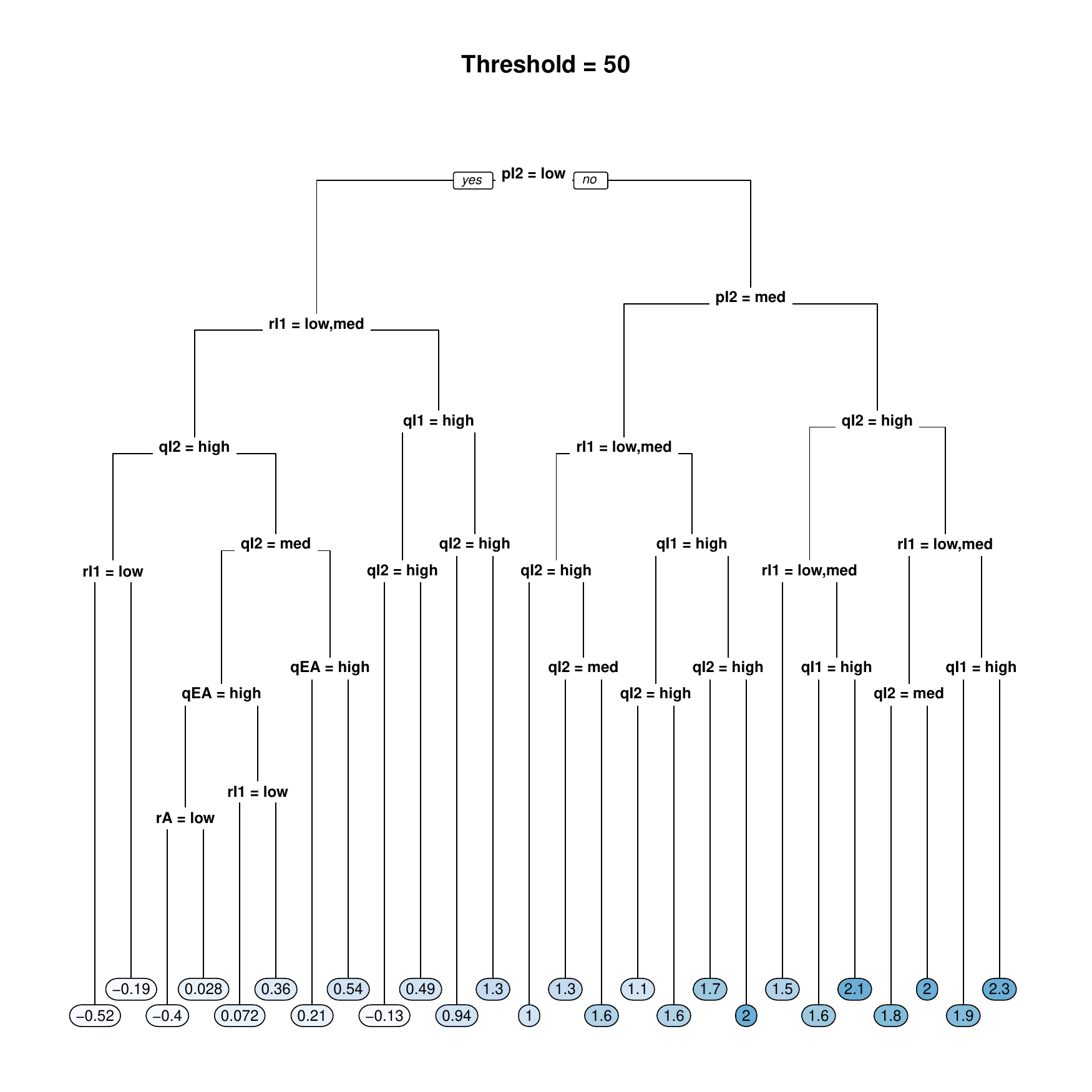}
	\caption{Pruned tree with 25 splits for predicting logit-CII with a class size threshold of 20. At each split, points which satisfy the listed condition move left, while the other points move right.}
	\label{fig:CII_splits_50}
\end{figure}

\begin{figure}[tp]
	\centering
	\includegraphics[width=\textwidth]{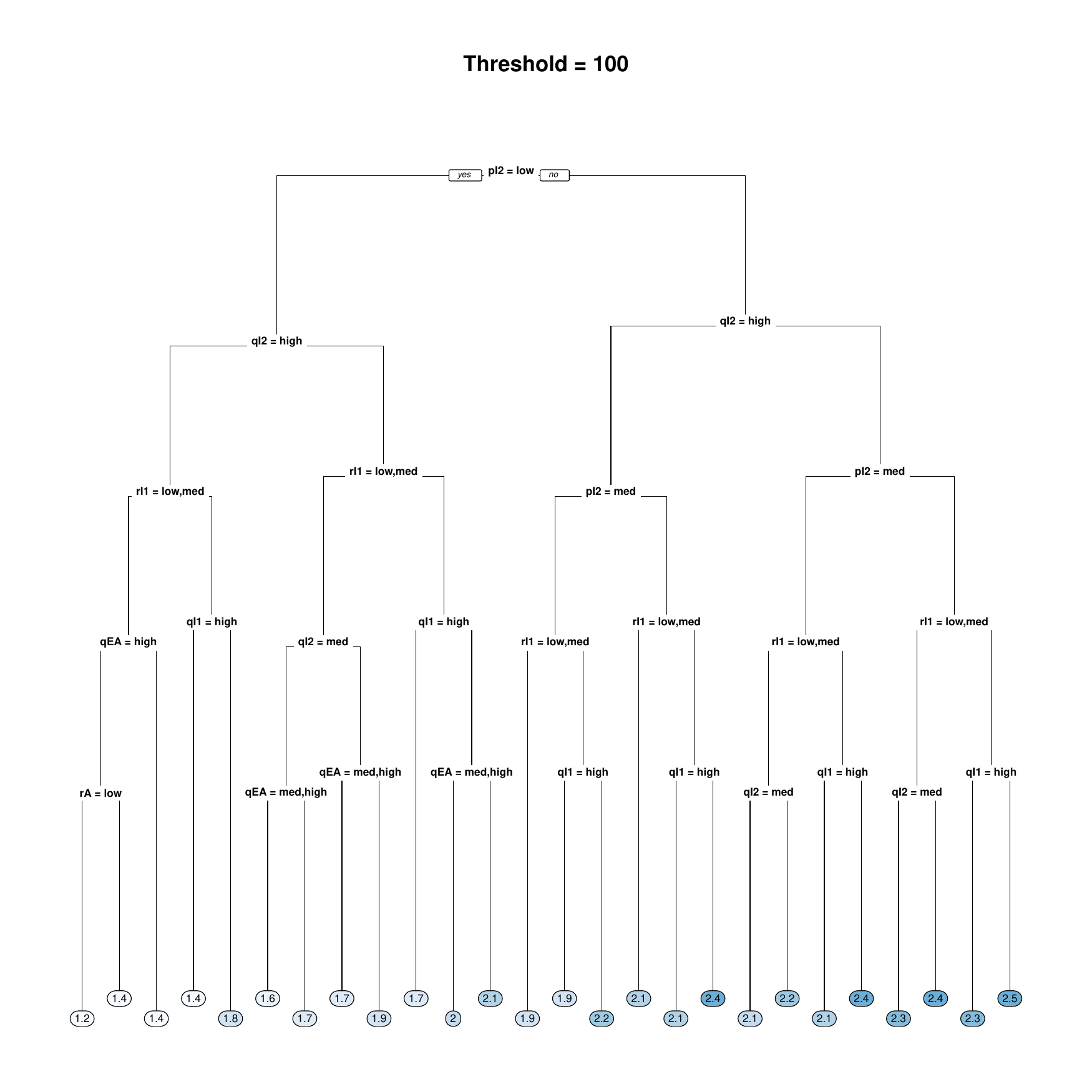}
	\caption{Pruned tree with 25 splits for predicting logit-CII with a class size threshold of 20. At each split, points which satisfy the listed condition move left, while the other points move right.}
	\label{fig:CII_splits_100}
\end{figure}

\begin{figure}[tp]
	\centering
	\includegraphics[width=\textwidth]{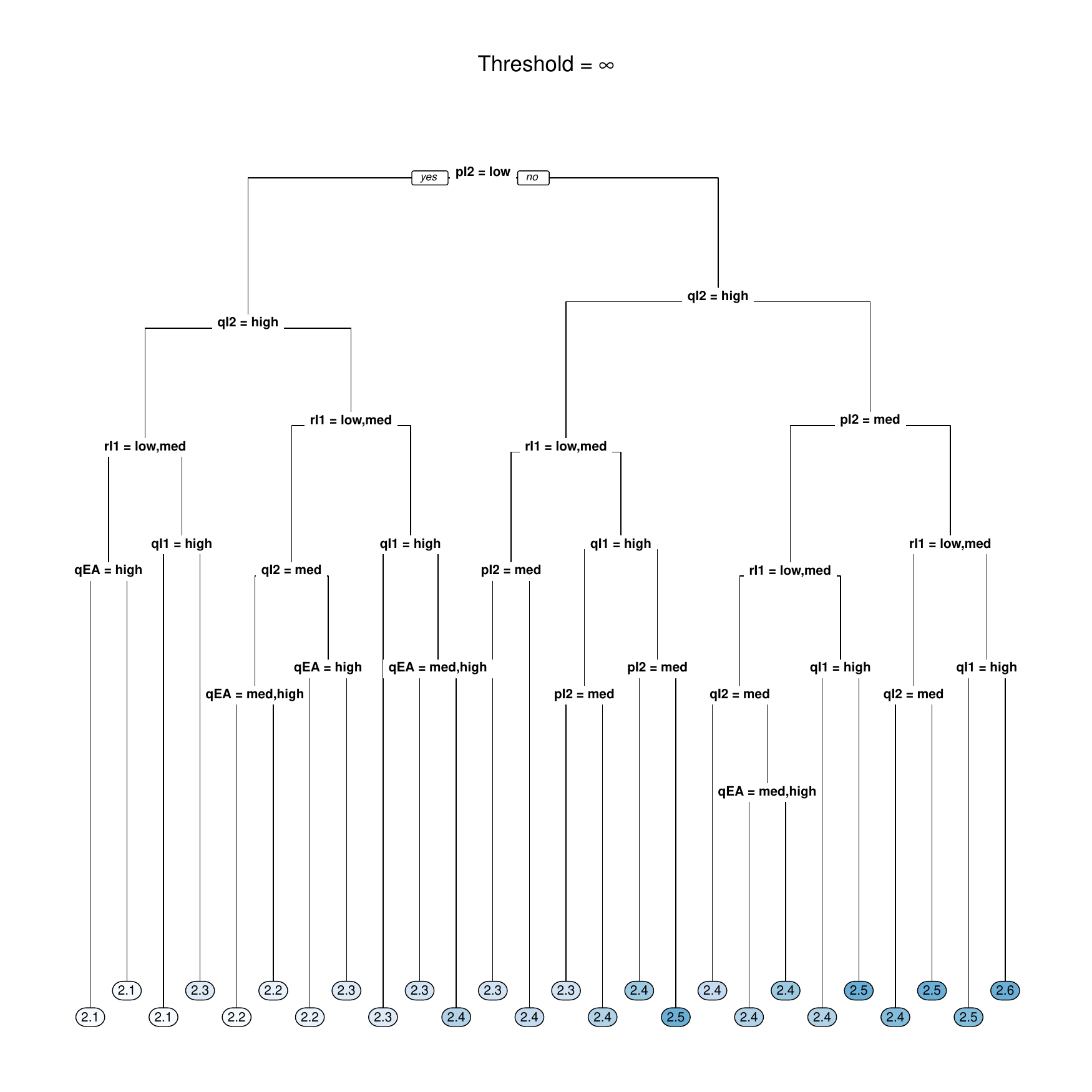}
	\caption{Pruned tree with 25 splits for predicting logit-CII with a class size threshold of 20. At each split, points which satisfy the listed condition move left, while the other points move right.}
	\label{fig:CII_splits_inf}
\end{figure}

\begin{figure}[tp]
	\centering
	\includegraphics[width=\textwidth]{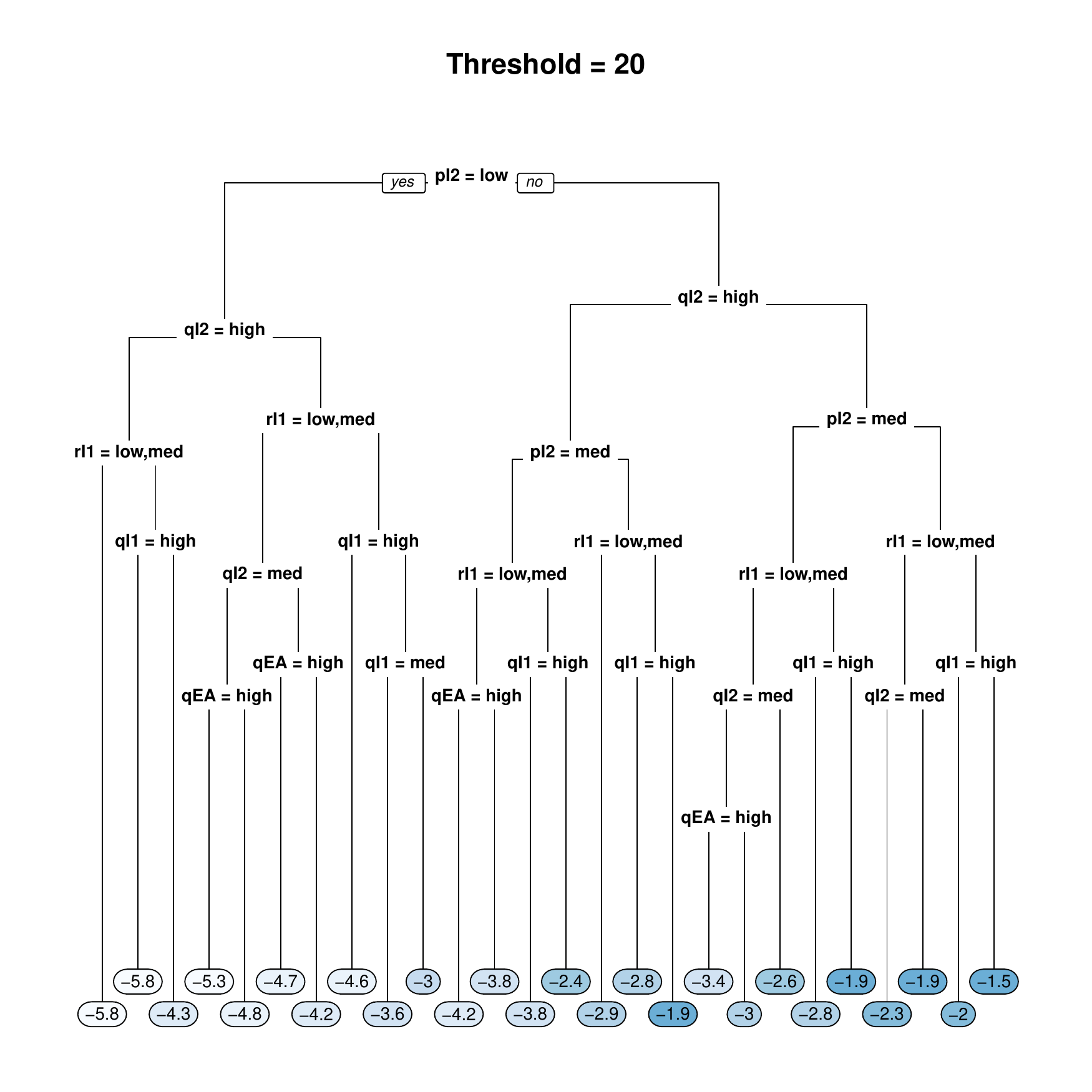}
	\caption{Pruned tree with 25 splits for predicting logit-peak outbreak size with a class size threshold of 20. At each split, points which satisfy the listed condition move left, while the other points move right.}
	\label{fig:peak_splits_20}
\end{figure}

\begin{figure}[tp]
	\centering
	\includegraphics[width=\textwidth]{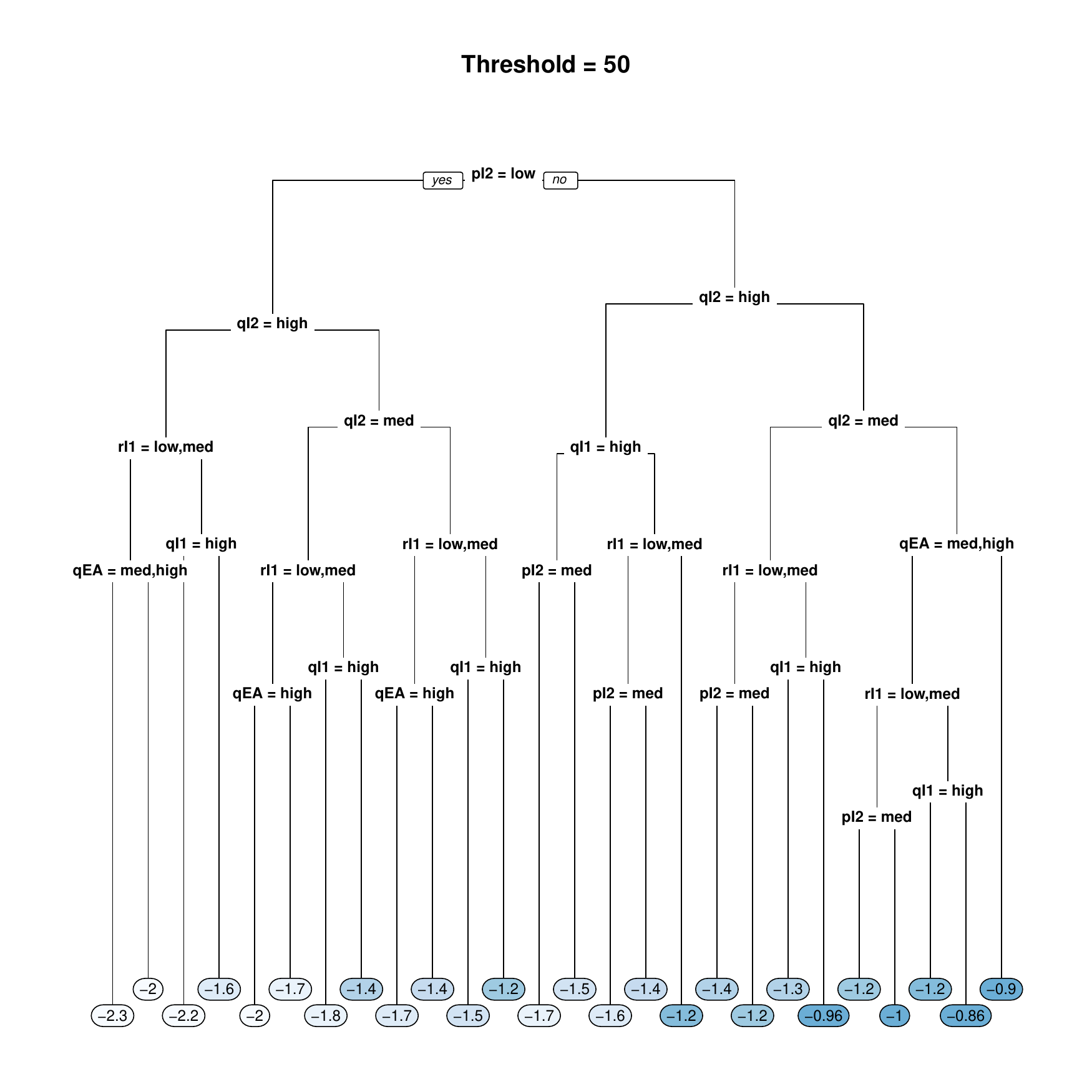}
	\caption{Pruned tree with 25 splits for predicting logit-peak outbreak size with a class size threshold of 20. At each split, points which satisfy the listed condition move left, while the other points move right.}
	\label{fig:peak_splits_50}
\end{figure}

\begin{figure}[tp]
	\centering
	\includegraphics[width=\textwidth]{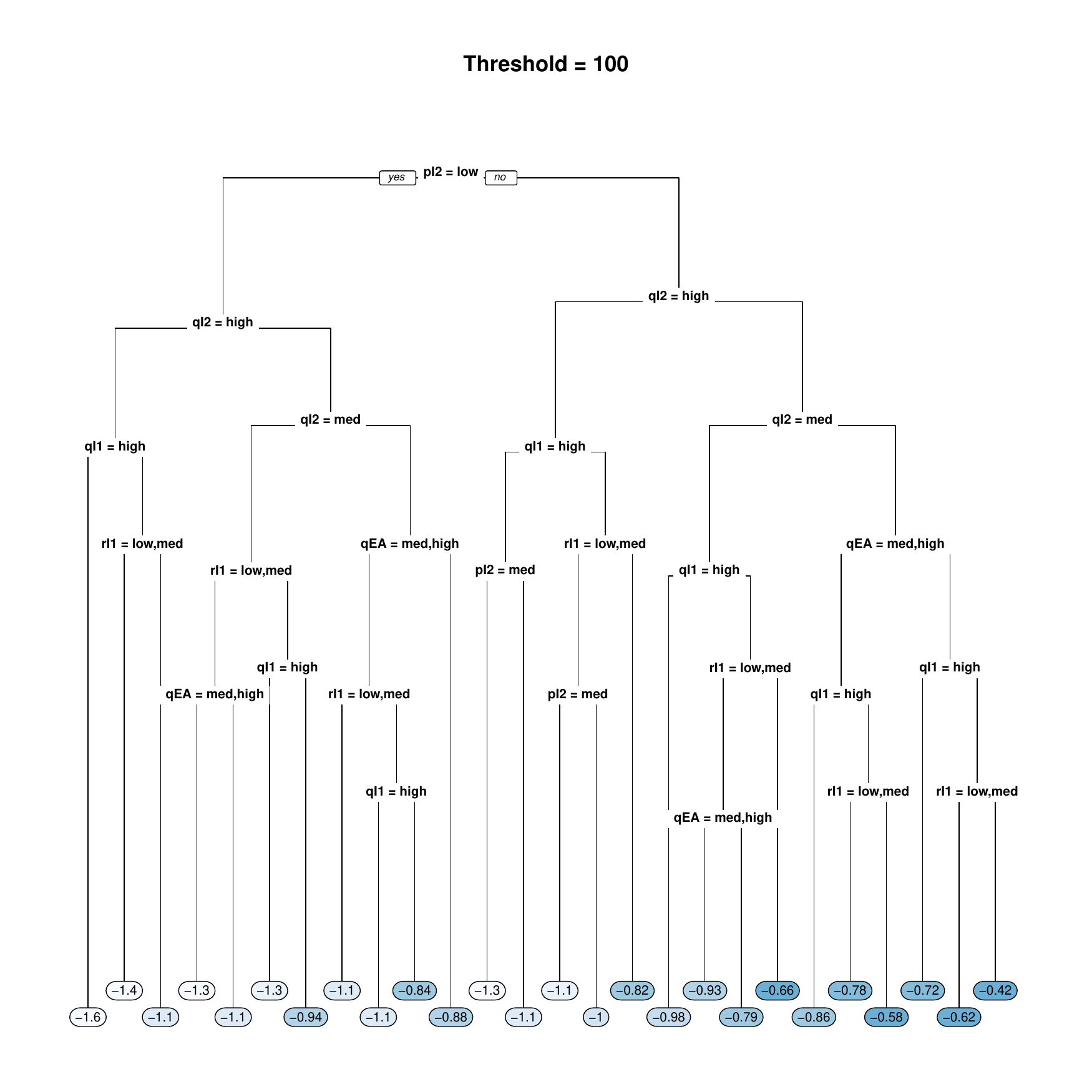}
	\caption{Pruned tree with 25 splits for predicting logit-peak outbreak size with a class size threshold of 20. At each split, points which satisfy the listed condition move left, while the other points move right.}
	\label{fig:peak_splits_100}
\end{figure}

\begin{figure}[tp]
	\centering
	\includegraphics[width=\textwidth]{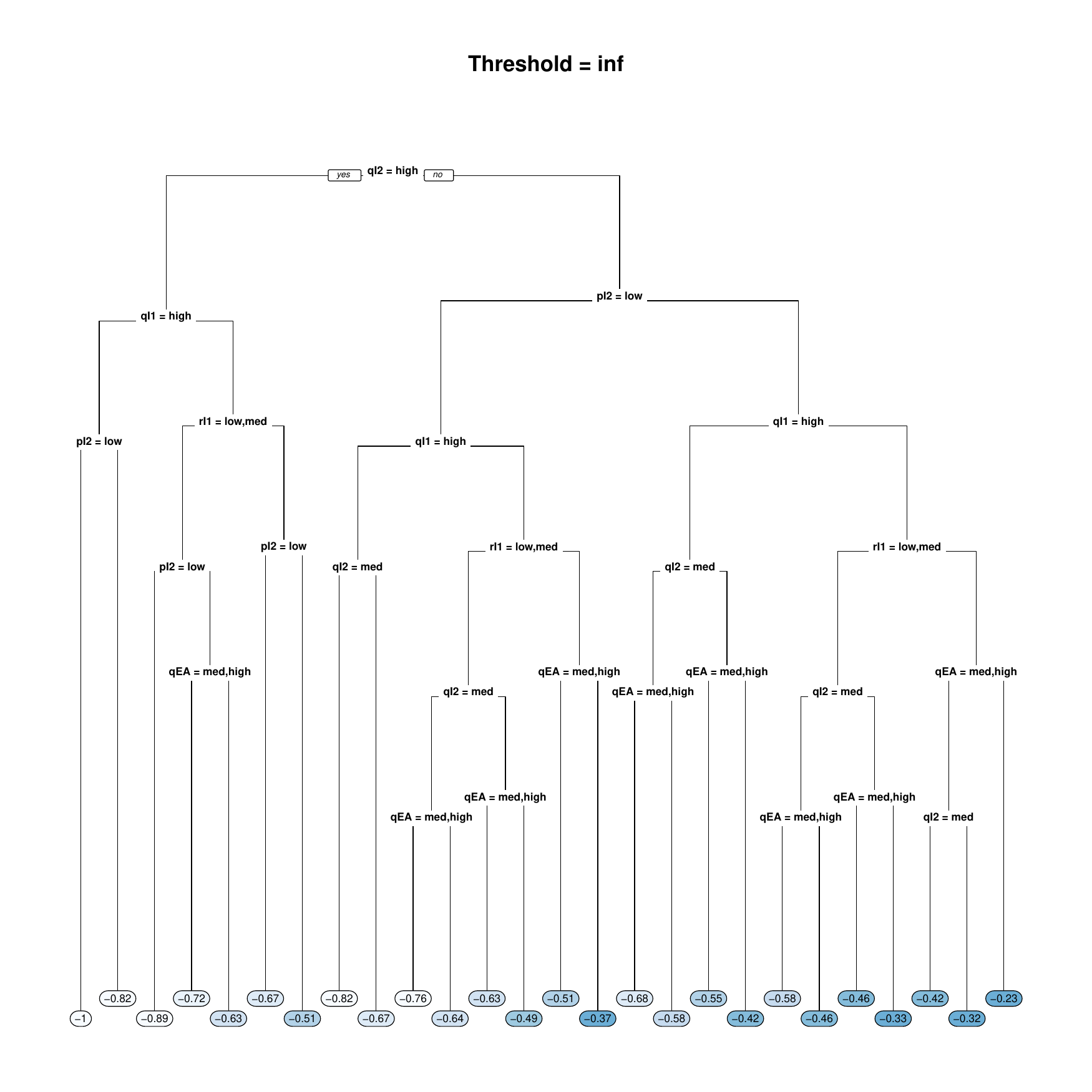}
	\caption{Pruned tree with 25 splits for predicting logit-peak outbreak size with a class size threshold of 20. At each split, points which satisfy the listed condition move left, while the other points move right.}
	\label{fig:peak_splits_inf}
\end{figure}

This appendix contains plots of the first 25 splits of regression trees fit to predict (logit-transformed) CII or peak outbreak size at each class size threshold. This gives a more qualitative description of the variable importance values in Tables \ref{tab:CII_var_imp} and \ref{tab:peak_var_imp}. Since these plots are produced in \texttt{R}, they use slightly different terminology for the parameters. Conversions are given in Table \ref{tab:appendix_par_names}. Figures \ref{fig:CII_splits_20}-\ref{fig:CII_splits_inf} give the CII trees, and Figures \ref{fig:peak_splits_20}-\ref{fig:peak_splits_inf} give the corresponding trees for predicting peak outbreak size.

\begin{table}[htp]
	\centering
	\begin{tabular}{c|c}
		Parameter & \texttt{R} Name\\
		\hline
		$\theta_{I2}$ & pI2\\
		$\rho_A$ & rA\\
		$\rho_{I1}$ & rI1\\
		$q_E$ & qE\\
		$q_A$ & qA\\
		$q_{I1}$ & qI1\\
		$q_{I2}$ & qI2\\
		$q_{EA}$ & qEA
	\end{tabular}
	\caption{Chosen values for each parameter, along with relevant references. See text for reasoning.}
	\label{tab:appendix_par_names}
\end{table}

\end{appendices}

\bibliographystyle{plainnat}
\bibliography{mybib}

\end{document}